\documentclass[aps,prapplied,twocolumn,byrevtex,showpacs,floatfix,superscriptaddress,reprint]{revtex4-2}

\usepackage[dvipsnames]{xcolor}
\usepackage{amsbsy}
\usepackage{amstext}
\usepackage[T1]{fontenc}
\usepackage[english]{babel}
\usepackage[utf8]{inputenc}
\usepackage{graphicx}
\usepackage{booktabs}
\usepackage{amsmath}
\usepackage{amsfonts}
\usepackage{amssymb}
\usepackage{float}
\usepackage{multirow}
\usepackage{colortbl}
\usepackage{appendix}
\usepackage[mathscr]{euscript}
\usepackage{physics}
\usepackage{bbold}
\usepackage{bm}
\usepackage{dsfont}

\usepackage{tikz}
\usetikzlibrary{quantikz}

\makeatletter
\newcommand{\dummylabel}[2]{\def\@currentlabel{#2}\label{#1}}
\makeatother

\usepackage{comment} 
\usepackage[linesnumbered,ruled,vlined]{algorithm2e} 

\newcounter{subroutine}
\newcounter{tmpalgocf}

\newenvironment{subroutine}{
        \SetArgSty{upshape}
        \setcounter{tmpalgocf}{\thealgocf}
        \setcounter{algocf}{\thesubroutine}
        \stepcounter{subroutine}
        \SetAlgorithmName{Subroutine}{subroutine}{List of Subroutines}
        \begin{algorithm}
}{
        \end{algorithm}
        \setcounter{algocf}{\thetmpalgocf}
}

\makeatletter

\usepackage{pifont}
\newcommand{\xmark}{\ding{54}}

\usepackage{hyperref}
\hypersetup{colorlinks, citecolor=NavyBlue, filecolor=black, linkcolor=black, urlcolor=black,
bookmarksnumbered=false, bookmarksopen=false}

\begin{document}
	\title{Hardware-efficient entangled measurements for variational quantum algorithms}
	\author{Francisco Escudero}
	\affiliation{Instituto de F{\'i}sica Fundamental, IFF-CSIC, Calle Serrano 113b, 28006 Madrid, Spain}
	\affiliation{CWI \& QuSoft, Science Park 123, 1098 XG Amsterdam, The Netherlands}
	\author{David Fern{\'a}ndez-Fern{\'a}ndez}
	\affiliation{Instituto de F{\'i}sica Fundamental, IFF-CSIC, Calle Serrano 113b, 28006 Madrid, Spain}
	\affiliation{Instituto de Ciencia de Materiales de Madrid, ICMM-CSIC, 28049 Madrid, Spain}
	\author{Gabriel Jaum{\`a}}
	\affiliation{Instituto de F{\'i}sica Fundamental, IFF-CSIC, Calle Serrano 113b, 28006 Madrid, Spain}
	\author{Guillermo F. Pe{\~n}as}
	\affiliation{Instituto de F{\'i}sica Fundamental, IFF-CSIC, Calle Serrano 113b, 28006 Madrid, Spain}
	\author{Luciano Pereira}
	\email[]{luciano.ivan@iff.csic.es}
	\affiliation{Instituto de F{\'i}sica Fundamental, IFF-CSIC, Calle Serrano 113b, 28006 Madrid, Spain}
	
	\begin{abstract}
	
        Variational algorithms have received significant attention in recent years due to their potential to solve practical problems using noisy intermediate-scale quantum (NISQ) devices. A fundamental step of these algorithms is the evaluation of the expected value of Hamiltonians, and hence efficient schemes to perform this task are required. The standard approach employs local measurements of Pauli operators and requires a large number of circuits. An alternative is to make use of entangled measurements, which might introduce additional gates between physically disconnected qubits that harm the performance. As a solution to this problem, we propose hardware-efficient entangled measurements (HEEM), that is, measurements that permit only entanglement between physically connected qubits. We show that this strategy enhances the evaluation of molecular Hamiltonians in NISQ devices by reducing the number of circuits required without increasing their depth. We provide quantitative metrics of how this approach offers better results than local measurements and arbitrarily entangled measurements. We estimate the ground-state energy of the H$_2$O molecule with classical simulators and quantum hardware using the variational quantum eigensolver with HEEM. 
	    
	\end{abstract}
	\maketitle
	
	\section{Introduction.}
	
    We are currently in the era of noisy intermediate-scale quantum computers (NISQ). The main limitations of these devices are short coherence times and noisy entanglement gates; therefore, NISQ circuits must inevitably have a low depth \cite{NISQ}. Given that, a lot of effort has been devoted to the design and implementation of quantum algorithms that only use low-depth circuits \cite{ Ryabinkin2018, Romero2018, Ryabinkin2018UCC, Kirby2021}. Among such algorithms, one family that has gained attention is variational quantum algorithms (VQAs) \cite{bharti2021, Tilly2021, Izmaylov2019}: hybrid quantum-classical methods where a classical computer guides a quantum computer to produce variational quantum states and to measure their expected value, with the goal of minimizing an objective function encoded in a Hamiltonian. The most famous VQAs are the variational quantum eigensolver (VQE) \cite{peruzzo2014} and the quantum approximate optimization algorithm (QAOA) \cite{farhi2014quantum}. Multiple VQAs have been applied in a wide range of areas, such as chemistry \cite{ Moll2018, kandala2017hardware, Hempel2018, Nam2020}, finance \cite{Egger2021, vikstal2020}, traffic prediction~\cite{Neukart2017}, machine learning \cite{Biamonte2017, Benedetti2019, Patterson2021, Chen2021_qml}, entanglement detection \cite{Wang2021, 2012.14311, 2110.03709}, and differential equations \cite{Yuan2019, Jones2019, 2104.02668}.
	
	Despite the great advances made in recent years, implementing VQA remains a challenge. One of the main drawbacks is the large number of measurements required to evaluate the objective function. Such evaluation can be done by decomposing the Hamiltonian on the basis of tensor products of Pauli operators, also called Pauli strings, and then measuring each of these terms independently. Using this approach, the number of measurements needed to evaluate the objective function is equal to the Pauli terms of the Hamiltonian, which scales as $N^4$ for typical instances such as second-quantized chemical Hamiltonians on $N$ qubits \cite{Jordan1928,Bravyi2002}. The preparation and measurement of each circuit require a non-negligible amount of time and resources; thus, decreasing the number of circuits is essential to speed up VQAs and to reach realistic applications of quantum computing.

	Several methods have been proposed to reduce the total number of measurements required to efficiently measure observables, such as classical shadows \cite{Huang2020,Chen2021}, quantum tomography \cite{Aaronson2017_qt, Cotler2020, Bonet-Monroig2020, Rubin2018}, machine learning \cite{Melko2019,Torlai2020}, sampling methods \cite{McClean2016, 2004.06252}, hamiltonian moments \cite{ Vallury2020}, and adaptive protocols \cite{2110.15339}. Other proposals are grouping methods, which exploit the commutative relation between the Pauli strings to make groups that can be measured simultaneously, reducing the total number of required experiments. The most widely known approach is the grouping with tensor product basis (TPB) \cite{bravyi2017, kandala2017hardware, Hempel2018, Nam2020, yen2020measuring}, which uses qubit-wise commutativity and does not require entanglement. Another alternative is to use entangled measurements, which further reduces the number of measurements compared to TPB \cite{Verteletskyi2019, gokhale2019minimizing, Kondo2022computationally,Izmaylov2020}. However, this last approach assumes unlimited entanglement resources and is not suitable for NISQ devices. There is a midpoint between the two extreme alternatives of assuming limitless entanglement or none at all: using measurements whose entanglement requirements are within the limits of NISQ devices. There are several works in this line \cite{hamamura2020efficient,crawford2021efficient,zhao2020measurement,jena2019pauli}, nevertheless, none of these techniques takes into account the connectivity of the particular quantum processor where the algorithm is run. 
	
	In this article, we address the problem of grouping Pauli strings with entangled measurements (EMs), but only between physically connected qubits, that is, hardware-efficient entangled measurements (HEEMs). Given a set of Pauli strings and the processor connectivity, there is a vast number of possible HEEMs, some of which will be more effective than others. Finding the optimal HEEM requires, among other things, finding the optimal mapping between the theoretical qubits of the algorithm and the physical qubits of the device. We have named this issue the \textit{processor embedding} problem. We propose heuristic algorithms to solve the \textit{processor embedding} problem and to perform the grouping with HEEMs. We run VQE with HEEM to estimate the ground-state energy of the H$_2$O molecule using both classical simulators and quantum hardware. The proposed method reduces the number of measurements due to the entangled measurements and avoids long-range qubit interactions that would involve noisy, deep circuits, thanks to the hardware-efficient approach.
	
    \section{Hardware-efficient grouping}\label{secGrouping}
	A general $N$-qubit Hamiltonian can be expanded in terms of Pauli strings as
	\begin{equation}
	    H =\sum_\alpha h_\alpha P_\alpha,
	    \label{eq:general_hamiltonian}
	\end{equation}
	where $P_\alpha$ are the tensor products of the identity and Pauli operators ($I,X,Y,Z$), and $h_\alpha \in \mathds{R}$ are the coefficients of each Pauli string. The standard routine to compute the expected value of $H$ consists of measuring each Pauli string $P_i$ with a TPB measurement, which are the tensor products of the bases of eigenstates of the Pauli operators $\{\mathcal{X},\mathcal{Y},\mathcal{Z}\}$, with $\mathcal{X} = \{\ket{0} \pm\ket{1}\}$, $\mathcal{Y} = \{\ket{0}\pm i\ket{1}\}$ and $\mathcal{Z} = \{\ket{0},\ket{1}\}$.

	Multiple Pauli strings can be evaluated simultaneously with a single TPB. In this case, we say that they are \textit{compatible} with that TPB. For example, the 3-qubit Pauli strings $XIZ$ and $XYZ$ are compatible with TPB $\mathcal{X} \otimes \mathcal{Y} \otimes \mathcal{Z}$. This property allows us to define a grouping method \cite{bravyi2017}, which consists of searching for the smallest set of TPB that can be used to evaluate the expected value of $H$. Finding the best TPB grouping for a given Hamiltonian is equivalent to finding the best coloring for its Pauli graph, which is NP-Complete \cite{ColoringSurvey, yen2020measuring}.
	
	One can go beyond the TPB grouping and measure the expected value of $H$ with fewer groups thanks to EM, that is, measuring after an entangling operation. In \cite{hamamura2020efficient} the authors proposed a heuristic algorithm to construct groups using EM between pairs of qubits (see Appendix~\ref{sec:ApenMeasurments} \cite{suppl}). The EM grouping proved to be more efficient than the TPB grouping, significantly reducing the number of measurements and the uncertainty in the evaluation of the observables. One drawback of this approach is that it neglects the error of entangling gates in NISQ devices, and hence it might actually worsen the results in some scenarios. This is particularly pronounced when performing entangling operations between nonphysically connected qubits, since these require the usage of mediating qubit---and hence additional entangling gates---that increase the depth and thus the error of the circuits.
	
	As a solution to this problem, we propose three heuristic algorithms to group Pauli strings with HEEM (see Appendix~\ref{algGrouping} \cite{suppl}). These algorithms introduce two important improvements to the algorithm proposed by Hamamura \textit{et al.} \cite{hamamura2020efficient}. First, the algorithms check whether two Pauli strings are simultaneously measurable by an EM given the connectivity of a device, isolating from all possible EM only those that are hardware efficient. Second, since the groups obtained by our algorithms or by Hamamura's algorithms depend on the orders followed by several loops, we introduce three algorithms to choose these orders. This is explained in the next section.
	
	\subsection{The processor embedding problem}\label{theophys}
    There are multiple factors that determine the performance of our grouping algorithms, for instance, the orders that its loops follow to run through the qubits, the assignable measurements, and the Pauli strings (here we refer to the orders of Subroutine~\ref{sub:2assignMeas} in Appendix~\ref{sec:ApendConnecitivyAlgorithms} \cite{suppl}). These orders can be optimized considering the \textit{processor embedding} problem: the problem of finding the optimal way to map the theoretical qubits of the Hamiltonian into the physical qubits of the device. In this section, we introduce a heuristic approach to solve this problem, which in turn allows us to optimize the order of the loops in the qubits and in the assignable measurements.

	Given the Hamiltonian shown in Eq.~(\ref{eq:general_hamiltonian}), we define its \textit{compatibility matrix} $C$ as an $N\times N$ symmetric matrix whose non-diagonal entries $C_{ij}$ are equal to the number of compatible entangled measurements (see Table~\ref{table1} in the Appendix~\ref{sec:ApenMeasurments} \cite{suppl}) involving qubits $i$ and $j$. Diagonal entries of $C$ are not relevant for the \textit{processor embedding} problem.
	
	The number of compatible entangled measurements that involve qubits $(i,j)$ is defined in the following way. The factors $i$ and $j$ of each Pauli string $P_\alpha$ can be regarded as a set of Pauli sub-strings of 2 qubits, $\{P^{ij}_\alpha\}$, and the number of compatible entangled measurements that involve qubits $(i,j)$ is the number of pairs of Pauli sub-strings $\{P_\alpha^{ij}\}$ that are compatible through an entangled measurement. Thus, the compatibility matrix encapsulates how many entangled measurements can be established between two qubits. As an example, consider the following set of Pauli strings
	\begin{equation*}
		P_\alpha\left\{ \begin{array}{l} XXZ \\ YYZ \\ YZZ \end{array}\right.
		\Rightarrow P_\alpha^{01}\left\{ \begin{array}{l} XX \\ YY \\ YZ \end{array}\right.\Rightarrow C_{01}=C_{10}=2.
	\end{equation*}
	In this example $C_{10}=C_{01}=2$ because $XX$ and $YY$ can be measured simultaneously with a Bell measurement and $XX$ and $YZ$ with an $\Omega^X$ measurement (see Appendix~\ref{Sumary} \cite{suppl}). However, $YY$ and $YZ$ cannot be measured simultaneously, i.e., $[YY, YZ] \neq 0$.
	
	\begin{figure}
		\centering
		\includegraphics[width=\linewidth]{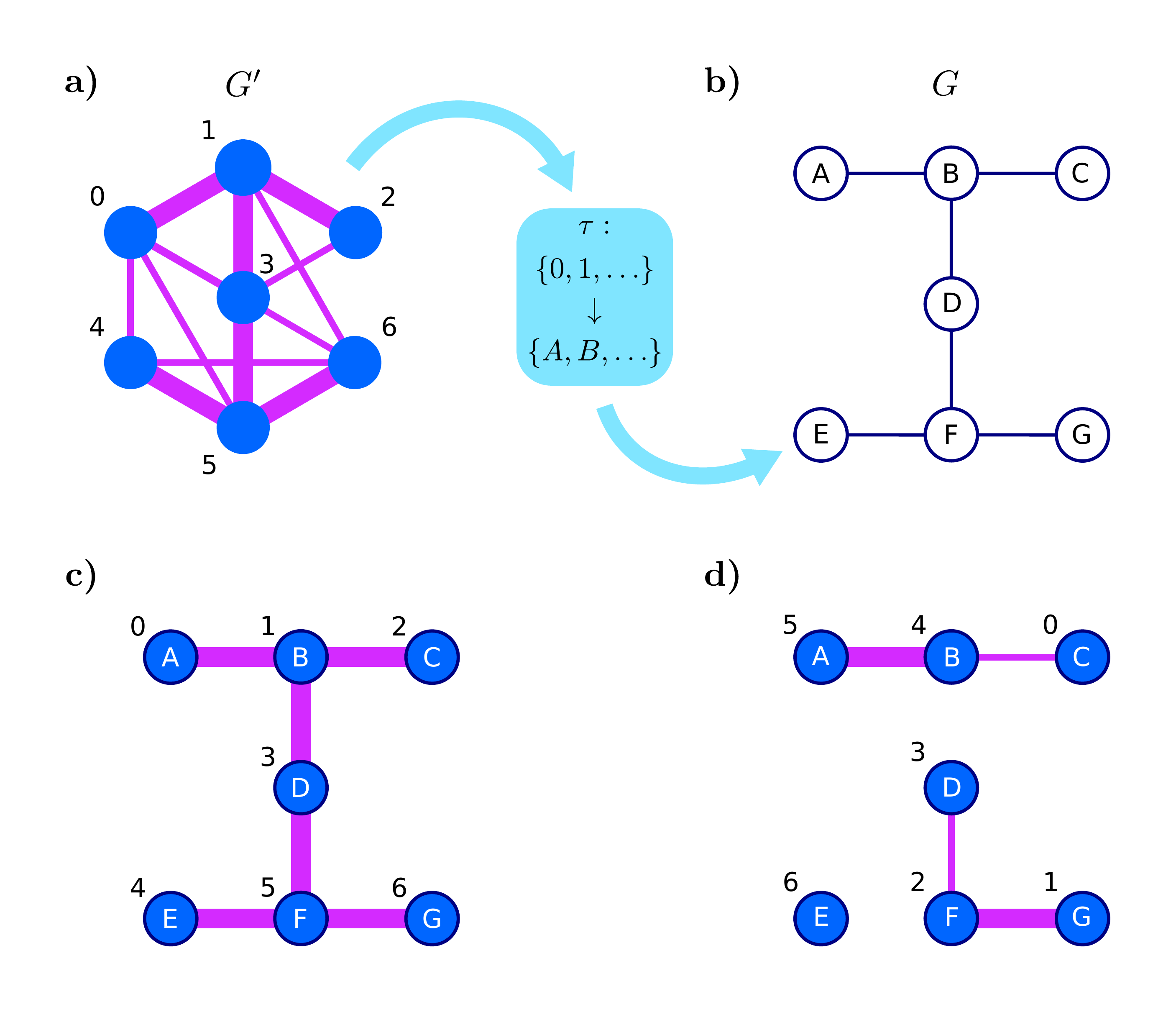}
		\caption{Schematic representation of the \textit{processor embedding} problem. \textbf{a)} Compatibility graph of theoretical qubits, $G'$. Given a set of Pauli strings, the width of the edges $\{i,j\}$ of this graph represents the number of entangled measurements that can be established between qubits $i$ and $j$, $C_{ij}$. \textbf{b)} Connectivity graph of the physical qubits of the \textit{ibmq\_jakarta} device, $G$. \textbf{c)} and \textbf{d)} are good and bad \textit{processor embedding} maps $\tau$, respectively.}
		\label{fig:scheme}
	\end{figure}
		
	With the compatibility matrix, we can tackle the \textit{processor embedding} problem by trying to maximize the number of potential entangled measurements between neighboring qubits. We formulate this problem in terms of graphs, as shown in Fig.~\ref{fig:scheme}. Let $G':=(V',E',C)$ be the weighted graph defined by the vertices $V':=\{0,\dots, N-1\}$, where $N$ is the number of theoretical qubits, the edges $E':=V'\times V'$ and the weights $C_{ij}$. In this way, $G'$ represents the compatibility between the theoretical qubits (Fig.~\ref{fig:scheme} a)). Let $G:=(V,E)$ be the graph defined by the vertices $V:=\{0,\dots, M-1\}$, with $M\geq N$ the number of physical qubits and the edges $E:=\{(i,j)\in V\times V:\  \mathrm{qubits}\ i\ \mathrm{and}\ j\ \mathrm{that\ are\ physically\ connected}\}$. In this way, $G$ represents the topology of the chip (Fig.~\ref{fig:scheme} b)). The objective is to find the map $\tau:\{0,\dots, N-1\}\to \{0,\dots, M-1\}$ that maximizes 
	\begin{equation}\label{eq:ObjectiveTheoPhys}
		\omega(\tau)=\sum_{(\tau(i),\tau(j))\in E} C_{ij}. 
	\end{equation}
	Note that $\omega(\tau)$ is the total number of compatibilities between theoretical qubits once they are mapped to physical qubits through $\tau$. For the example shown in Fig.~\ref{fig:scheme} a), where the thin edges correspond to $C_{ij}=1$ and the wide ones to $C_{ij}=2$, the optimal \textit{processor embedding} map (Fig.~\ref{fig:scheme} c)) results in $\omega(\tau)=12$, while a bad mapping (Fig.~\ref{fig:scheme} d)) gives a lower value $\omega(\tau)=6$.
 
\begin{figure}
		\centering
		\includegraphics[width=\linewidth]{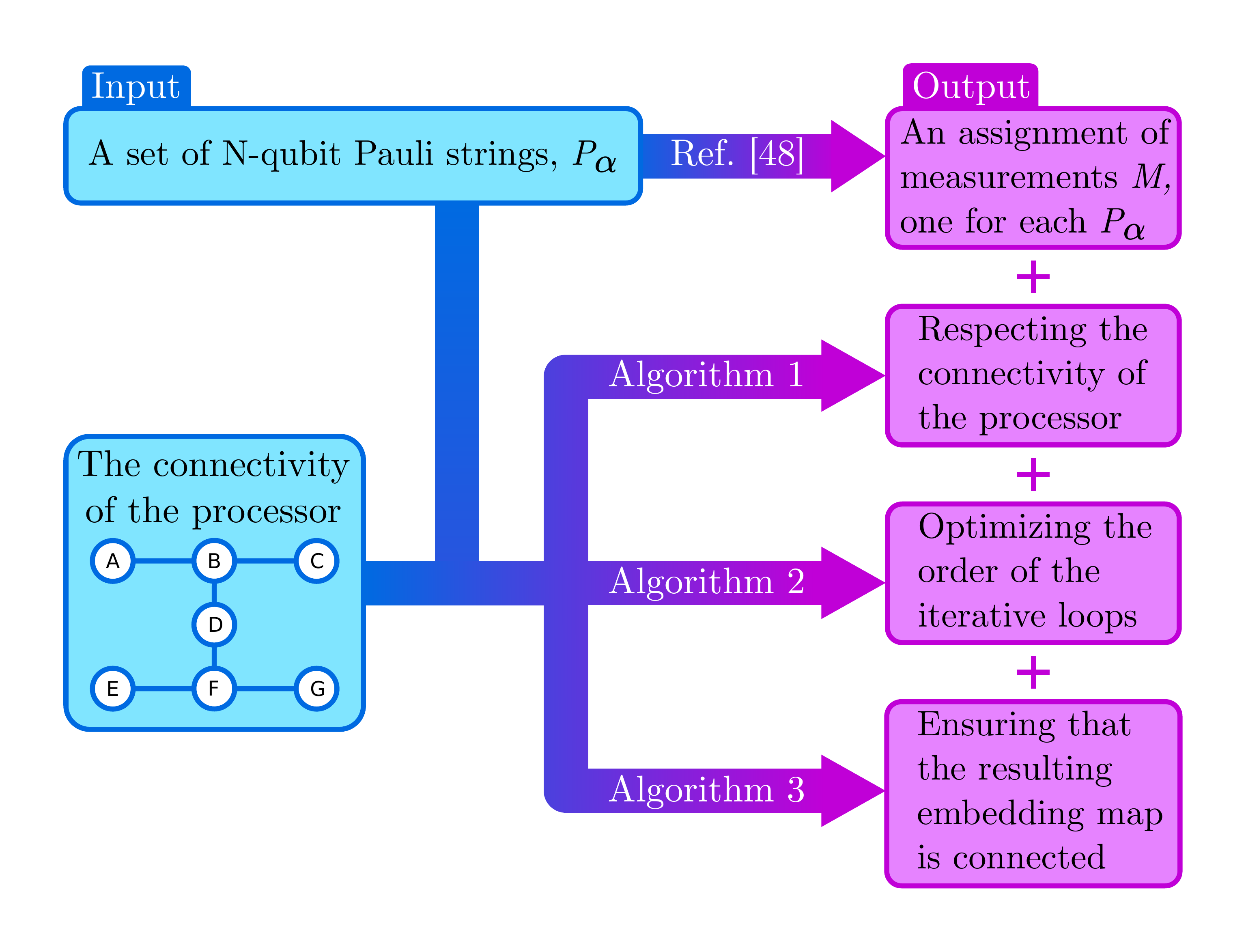}
		\caption{Flow chart of the different algorithms used in this work. The algorithm shown in \cite{hamamura2020efficient} is referred in the main text as EM. The Algorithms~\ref{alg:NaiveGrouping'}, \ref{Alg:OrdDisc'} and \ref{Alg:OrdConnected'} are named \emph{naive}, \emph{order-disconnected}, and \emph{order-connected}, respectively. These three algorithms are gathered under HEEM grouping.}
		\label{fig:flujo}
	\end{figure}

    The problem of finding the best $\tau$ is what we refer to by the \textit{processor embedding} problem, which we know is NP-Hard, and hence there is no efficient algorithm for the general case. There are several ways to prove that the \textit{processor embedding} problem is NP-Hard. For example, the max-cliqué problem, which is itself NP-Hard \cite{Karp1972}, can be regarded as a simplification of the \textit{processor embedding} problem so that all nondiagonal entries of $C$ are positive and equal. Additionally, the k-densest subgraph problem is NP-Complete and can also be seen as a particular instance of the \textit{processor embedding} problem \cite{Manurangsi2016, Sotirov2019}. Efficient algorithms could be found for the physically relevant instances of the \textit{processor embedding} problem if one finds some kind of structure within them. For now, we propose two heuristic algorithms (see Appendix~\ref{sec:AlgOrd} \cite{suppl}) to construct the map $\tau$. The first is the \textit{order-disconnected} map (Algorithm~\ref{Alg:OrdDisc'} and Subroutine~\ref{sub:3theophysNoncon}). Here, each pair of theoretical qubits $(i,j)$ with the highest entries $C_{ij}$ is assigned to physically connected qubits. The second alternative is the \textit{order-connected} map (Algorithm~\ref{Alg:OrdConnected'} and Subroutine~\ref{sub:4theophysConnected}). This algorithm does the same as the previous, but it additionally ensures that the graph $\tau(G')$ is a connected sub-graph of $G$. The third alternative is the \textit{naive} map (Algorithm~\ref{alg:NaiveGrouping'}). This is simply the trivial map $\tau(i)=i$ for $i\in \{0,\dots,N-1\}$, and it allows us to benchmark our \textit{order-connected} and \textit{order-disconnected} maps. An schematic flow chart of the different algorithms is shown in Fig.~\ref{fig:flujo}.

    Once the map $\tau$ has been chosen, one can propose an order for the loops of our grouping algorithm (Subroutine~\ref{sub:2assignMeas}), both in qubits and measurements. For this purpose, we introduce the $\tau$-compatibility matrix $C^\tau$
    \begin{equation}
        C_{ij}^\tau:=\left\{\begin{array}{lll} C_{ij} & \mathrm{if} & (\tau(i),\tau(j))\in E\\ 0 & \mathrm{if} & (\tau(i),\tau(j))\notin E. \end{array}\right.
    \end{equation}
     
	\noindent Let ${CQ}^\tau$ be the $N$-vector whose $i$ entry is given by 
	\begin{equation}
		{CQ}^\tau_i:={C}\mathcal{X}_i+{C}\mathcal{Y}_i+{C}\mathcal{Z}_i+\sum_{j\neq i}{C}^\tau_{ij},
	\end{equation}
	where ${C}\mathcal{X}_i$, ${C}\mathcal{Y}_i$, and ${C}\mathcal{Z}_i$ are the numbers of compatibilities involving the qubit $i$ through measurements $\mathcal{X}$, $\mathcal{Y}$, and $\mathcal{Z}$, respectively. In this way, ${CQ}^\tau_i$ is the number of compatibilities involving the qubit $i$, since $\tau$ has been chosen as the embedding map of the processor. Now, we make the natural choice of running through the qubits in descending order of ${CQ}^\tau.$
	
\begin{figure*}[ht!]
	\centering
	\includegraphics[width=\linewidth]{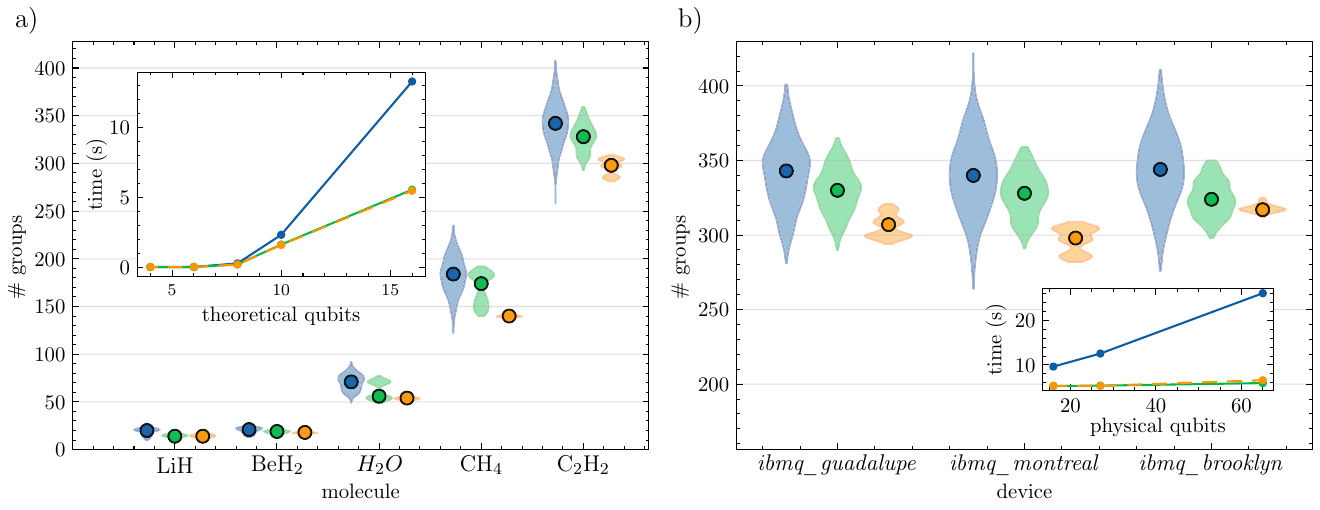}
	\caption{Results for grouping algorithms. The violins correspond to the distributions on the number of groups after running a Monte Carlo method over the order of the Pauli strings and the order of the theoretical qubits for different \textbf{a)} molecules, and \textbf{b)} devices. Blue violins correspond to the \emph{naive} algorithm (see Appendix~\ref{algGrouping} Algorithm~\ref{alg:NaiveGrouping'} \cite{suppl}), green violins correspond to the \emph{order-disconnected} algorithm (Algorithm~\ref{Alg:OrdDisc'}), and orange violins to the \emph{order-connected} algorithm (Algorithm~\ref{Alg:OrdConnected'}). The colored dots represent the mean value in the number of groups in each case. The insets show the average execution time of the grouping algorithm. In \textbf{a)} the algorithms are grouped according to HEEM with the connectivity of \textit{ibmq\_montreal}. In \textbf{b)} the number of groups corresponds to the C$_2$H$_2$ molecule.}
	\label{fig:violinHEEM}
\end{figure*}

    Let $CM^\tau$ be the 9-vector given by 
    \begin{multline}
		CM^\tau:=(CM_\mathcal{X},CM_\mathcal{Y},CM_\mathcal{Z},CM^\tau_\mathrm{Bell},\\ CM^\tau_{\Omega^X}, CM^\tau_{\Omega^Y},CM^\tau_{\Omega^Z},CM^\tau_\chi,CM^\tau_{\tilde{\chi}}),
	\end{multline}
    where $CM^\tau_\mathrm{Bell}$ is the number of compatibilities due to Bell measurement once $\tau$ has been chosen as the \textit{processor embedding} map. $CM^\tau_{\Omega^X}$, $CM^\tau_{\Omega^Y}$, $C^\tau_{\Omega^Z}$, $CM^\tau_\chi$ and $CM^\tau_{\tilde{\chi}} $ are defined analogously for the other entangled measurements (see Appendix~\ref{sec:ApenMeasurments} \cite{suppl}). $CM_\mathcal{X}$, $CM_\mathcal{Y}$ and $CM_\mathcal{Z}$ are the numbers of compatibilities by the measurements $\mathcal{X}$, $\mathcal{Y}$ and $\mathcal{Z}$ (these numbers do not depend on $\tau$ because these measurements are separable). Now, similarly to what we did for the order of the qubits, we choose to run through the measurements in descending order of $CM^\tau$.
    
    An additional optimization that remains is the iterative order in which the algorithms visit each of the Pauli strings. As the \textit{processor embedding} map does not provide means to optimize this order, we adopt the traditional approach. This consists of building the Pauli graph and visiting the Pauli strings in descending order with respect to their degree in this graph, similar to what is done in the largest degree-first coloring algorithm (LDFC) \cite{LDFC}.
	
    Finally, we must emphasize that the \textit{processor embedding} problem is not only relevant in choosing the best orders for the loops of our grouping algorithm (Subroutine~\ref{sub:2assignMeas}), but also affects the number of groups obtained. For example, suppose that we have $N=M=3$, two Pauli strings $XXZ$ and $ZXX$ and $E=\{(0,1),(1,2)\}$.  If we choose $\tau:\ 0\to 0,\ 1\to 1,\ 2\to 2$ we need to measure both strings separately, but choosing $\tilde{\tau}:\ 0\to 1,\ 1\to 0,\ 2\to 2$ a single measurement suffices. In addition, the \textit{processor embedding} problem is crucial to reduce the error caused by the gates on NISQ computers, as a good choice of $\tau$ would also reduce the number of CNOTs used in the algorithm that precedes the measurement, which is a relevant task in the physical layout problem \cite{tan2020optimal}. In fact, if we had to apply a theoretical CNOT in our algorithm between qubits $i$ and $j$ that are mapped to physical qubits whose distance on the chip is $D$, then in practice we would need to apply $\order{D}$ CNOTs between physically connected qubits, so reducing $D$ using an appropriate $\tau$ is essential. Also, in a real device, the accuracy of the CNOT gates between connected qubits will vary depending on the chosen qubits. If one takes into account these other tasks and not only the grouping, the weights of $G'$ should vary: it should depend not only on the compatibility matrix, but also on those other features. This extension that takes into account both the quantum algorithm and the measurement scheme will be addressed in future works.

\section{Results}

\begin{table*}[ht!]
    	\centering
    	\resizebox{1.9\columnwidth}{!}{
    		\begin{tabular}{c|c|c|c|c|c|c|c|c|c|c|}
    			\cline{2-11}
    			\multirow{2}{*}{} & \multirow{2}{*}{Qubits} & \multirow{2}{*}{No grouping} & 
    			\multicolumn{3}{c|}{Grouping} & \multicolumn{2}{c|}{CNOTs} & \multicolumn{3}{c|}{Relative error (\%)} \\ 
    			\cline{4-11}
    			&  & & TPB & EM & HEEM & EM & HEEM & TPB & EM & HEEM \\ \hline
    			\multicolumn{1}{|l|}{H$_2$} & 2 & 5 & 2 & 2 & 2 & 1 & 1 & $2.9\pm0.3$& $2.5\pm0.2$& $2.4\pm0.3$ \\ \hline
    			\multicolumn{1}{|l|}{LiH} & 4 & 100 & 25&11&10 & 8&8 & $0\pm1$&$0\pm1$&$0.2\pm0.9$ \\ \hline
    			\multicolumn{1}{|l|}{BeH$_2$} & 6 & 95 & 24 & 15 & 13 & 74 & 18 &-&-&- \\ \hline
    			\multicolumn{1}{|l|}{H$_2$O} & 8 & 444 & 93 & 51 & 47 & 563 & 80 &$7\pm1$&$9\pm1$&$3.2\pm0.8$ \\ \hline
    			\multicolumn{1}{|l|}{CH$_4$} & 10 & 1181 & 246 & 113 &117& 2677 & 224 &$15\pm2$&$11\pm2$&$6\pm3$  \\ \hline
    			\multicolumn{1}{|l|}{C$_2$H$_2$} & 16 & 1884 & 457 & 189 & 258& 8969& 433&$16\pm2$&$19\pm2$&$12\pm3$ \\ \hline
    			\multicolumn{1}{|l|}{CH$_3$OH} & 22& 9257& 2225& 682& 1503& 9830& 2770& $25\pm5$& $31\pm3$& $20\pm4$\\ \hline
    			\multicolumn{1}{|l|}{C$_2$H$_6$} & 26& 8919& 2069& 758& 1529& 55809& 2873& $34\pm7$& $40\pm3$& $22\pm4$\\ \hline
    		\end{tabular}
    	}
    \caption{Number of groups, number of CNOTs, and relative error of the energy evaluation for some molecules using different grouping strategies. TPB groups have been obtained using the LDFC algorithm. EM groups are those proposed by \cite{hamamura2020efficient}. HEEM groups have been obtained using the best performing method among Algorithms~\ref{alg:NaiveGrouping'}, \ref{Alg:OrdDisc'} and \ref{Alg:OrdConnected'} in each case, and assuming that the physical qubits have the connectivity of the device \textit{ibmq\_montreal}. The relative error refers to the error in the estimation of energy for an initial state $\ket{0}^{\otimes N}$. Simulations are carried out considering the noise model of the device \textit{ibmq\_montreal}, and the uncertainty in relative error is given by the standard deviation for a total of 25 simulations. Each simulation has a total of $2^{14}$ shots evenly distributed across all measurements in each grouping. We omit the relative error of BeH$_2$ because the expected energy is equal to zero.}
    \label{table:groupings}
\end{table*}

    We begin by comparing the three different HEEM grouping algorithms that we have proposed: the \emph{naive}, the \emph{order-disconnected}, and the \emph{order-connected} methods (Algorithms~\ref{alg:NaiveGrouping'}, \ref{Alg:OrdDisc'} and \ref{Alg:OrdConnected'}). The reasons that led us to develop the \emph{order-connected} and \emph{order-disconnected} methods were improving the \emph{naive} method and reducing its dependence on the order followed in its loops. To verify the enhancements, we studied the performance of the three algorithms in several molecular Hamiltonians. They were constructed from their fermionic counterpart on STO3G basis \cite{Young2001} with the parity map \cite{McArdle2020}. In addition, we freeze the core and remove the unoccupied orbital of the molecules to reduce their number of qubits (for more details, see \cite{github}).
    
    Despite the optimizations implemented in the algorithms, their effectiveness still depends on the initial order of the theoretical qubits and the Pauli strings in the Hamiltonian. To address this dependency, we employed a Monte Carlo method to evaluate the average performance of the three algorithms. For a given Hamiltonian, we consider the set of all permutations of the order of the theoretical qubits and of the Pauli strings of that specific Hamiltonian. Each element of this set represents a potential input for the grouping algorithms. Monte Carlo is executed by randomly selecting elements on this set. Figure~\ref{fig:violinHEEM}~a) shows that the \emph{order-connected} and \emph{order-disconnected} algorithms are better on average than the \emph{naive} algorithm and that they have less dispersion, reducing their dependence on the orders. One might note that the absolute minimum number of groups is obtained with the \emph{naive} algorithm, however, this is not relevant in practice since one would like to run the algorithm just once. It is also remarkable that the \emph{order-connected} algorithm is the one that performs best by having both the lowest average value and the smallest dispersion, which means that this algorithm explores a region of the HEEM space that is more densely packed with good solutions.
    
    In terms of execution time, the \emph{order-connected} and \emph{order-disconnected} algorithms are much faster than the \emph{naive} one. The time complexity of the \emph{order-connected} and \emph{disconnected} algorithms is $\order{N^{5}}$ on the number of qubits in the Hamiltonian, versus the time complexity $\order{N^{5.5}}$ needed for a \emph{naive} grouping.  Figure~\ref{fig:violinHEEM} b) compares the performance of these three algorithms with three different processor architectures. It shows that the execution time of the non-naive algorithms does not depend on the architecture in contrast to the \emph{naive} one. This result is important for the practical application of the proposed algorithms on quantum devices with a large number of qubits. The opposite situation occurs for the grouping configurations: the \emph{naive} algorithm does not differentiate between processors, while the other two do. Thus, only the non-naive algorithms have the desired behavior with respect to different architectures: they are always quickly executed and can take advantage of the topology of the chip.

    Let us now confront our grouping methods with the previous ones. Table~\ref{table:groupings} shows the number of groups obtained from Pauli strings of molecules of sizes between 2 and 26 qubits using different methods: TPB grouping \cite{bravyi2017}, EM grouping \cite{hamamura2020efficient}, and HEEM grouping with the connectivity of the quantum device \textit{ibmq\_montreal}. Analogously to the results of Fig.~\ref{fig:violinHEEM}, the Hamiltonians were constructed with the parity map on the basis of STO3G, freezing the core and removing unoccupied orbitals \cite{github}. The results of the HEEM grouping correspond to the minimum number of groups obtained with our Monte Carlo study. Table~\ref{table:groupings} also includes the number of CNOTs required to measure the groups. On the one hand, the HEEM grouping significantly outperforms TPB in terms of the number of groups. However, the EM grouping has a smaller number of groups than the HEEM but uses a number of CNOTs that grows much more rapidly with the size of the molecules. The number of groups with HEEM scales as $\order{N^{2.45}}$ with the number of qubits. Furthermore, HEEM circuits involve only one layer of entangling gates, while the depth of EM circuits depends on the distance between the qubits to be connected. All the scalings shown here were obtained by fitting the numerical results obtained with each grouping. More details on this topic, together with the asymptotic scaling values for other magnitudes, can be found in the Appendix~\ref{SMsec:scaling} \cite{suppl}.

    Both the number of CNOTs, which translates into experimental error, and the number of groups, which translates into statistical error for a fixed total number of shots, are relevant for the final accuracy of the method. This means that the results discussed so far do not completely conclude which is the best grouping method. In Table~\ref{table:groupings} we show the relative error for the energy evaluation of the state $\ket{\psi} = \ket{0}^{\otimes N}$ for different molecule Hamiltonians and grouping methods. We consider this state because, having no entanglement, it is prepared with high accuracy. This implies that in the simulations only the error due to the energy evaluation appears. We set the total number of shots at $2^{14}$ for all molecules and methods, and these shots are divided equally between all circuits of each scenario. The relative error is computed as
    \begin{equation}
        \mathrm{Relative\; error} = \abs{(E_\mathrm{exact} - E_\mathrm{simulation})/E_\mathrm{exact}},
    \end{equation}
    where the exact energy is obtained as $E_\mathrm{exact}=\bra{\psi}H\ket{\psi}$, while $E_\mathrm{simulation}$ is calculated considering the noise model of the device \textit{ibmq\_montreal}. This simulation is repeated 25 times and from there we compute the average relative error shown in Table~\ref{table:groupings}. The uncertainty in this relative error is defined by the standard deviation between the different runs.
  
    With TPB grouping, no entangled measurements are performed, and hence there are no errors due to two-qubit gates. However, the total number of circuits to compute the energy is the largest, and consequently, TPB has the lowest number of shots per circuit and the highest statistical error. With EM, the scenario is the opposite; this method has the lowest statistical error, but the highest experimental error due to the large number of CNOT gates used to connect distant qubits. HEEM takes the best of both methods, as it uses fewer groups than TPB, reducing the statistical error, and it does not connect distant qubits, reducing the experimental error. Because of this, as can be seen in Table~\ref{table:groupings}, HEEM provides the lowest relative error for all the studied molecules.

\begin{figure}[ht!]
		\centering
		\includegraphics[width=\linewidth]{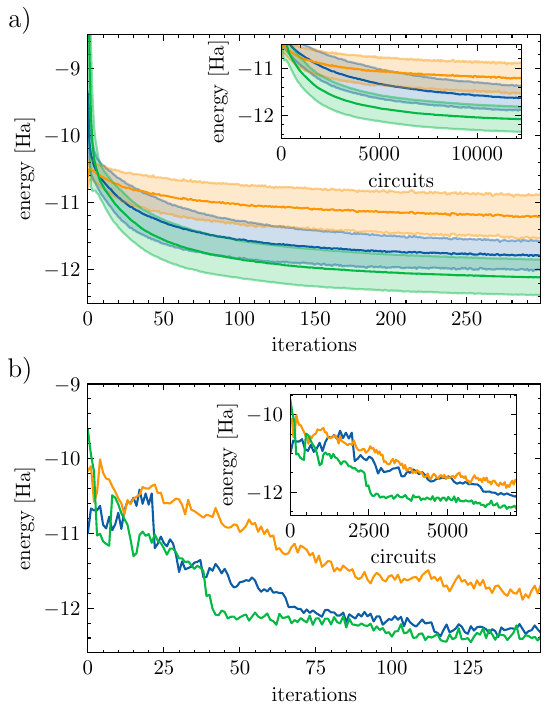}
		\caption{Simulation \textbf{a)}, and experimental implementation \textbf{b)}, of the VQE for the 8-qubits ${\rm H_2O}$ Hamiltonian, with a distance of $d=0.96$~\r{A} between hydrogen and oxygen atoms, for TPB in blue, EM in orange, and HEEM in green lines. The simulations were performed considering the basis gate, noise model, and connectivity of \textit{ibmq\_montreal}, while the real experiment is performed in \textit{ibmq\_guadalupe}. The classical optimizer used was SPSA with 300 iterations for the simulations and 150 iterations for the experiment. There are $2^{14}$ shots per circuit in both cases. \textbf{a)} Each point represents 240 independent instances of VQE where the solid lines correspond to the mean energy and the shaded region to the standard deviation. In both panels, the insets show the mean energy versus the number of circuits needed to measure the Hamiltonian.}
		\label{fig:VQE_result}
	\end{figure}

    Finally, to study the performance of the HEEM grouping in a practical scenario, we have run noisy simulations and a NISQ experiment of the VQE for the H$_2$O molecule at its bond distance. The Hamiltonian was constructed by the parity map from the fermionic Hamiltonian on STO3G basis. In addition, we freeze the core and remove unoccupied orbitals of the H$_2$O molecule to reduce the number of theoretical qubits to 8, so that the minimum energy is given by $-13.9$~Ha. Figure~\ref{fig:VQE_result} a) shows the mean energy per iteration of 240 instances of noisy VQE simulations using TPB, EM and HEEM. They were carried out using the noise model and the connectivity of \textit{ibmq\_montreal}, provided by IBM-Q \cite{IBMQ}. The variational ansatz is composed of 2 layers of local gates and 1 layer of CNOT between the physically connected qubits (see Appendix~\ref{variational_circuit} \cite{suppl}). The minimum energy attainable with this ansatz is $-12.3$~Ha. The groups used for HEEM correspond to the best groups obtained among the three proposed algorithms. We can see that the better performance grouping scheme is HEEM, achieving energies below $-12.0$~Ha after 200 iterations, while neither TPB nor EM reaches this value in 300 iterations. 
    The inset of Fig.~\ref{fig:VQE_result} a) shows the mean energy in terms of the number of circuits. We can see that the HEEM grouping achieves energy below $-12.0$~Ha with about $1200$ circuits, while TPB and EM only attain $-11.6$~Ha and $-11.2$~Ha, respectively. This means that HEEM allows us to obtain the same results as the previous approaches, but with fewer circuits.

    We also performed an experimental implementation of the VQE for the H$_2$O molecule with TPB, EM, and HEEM in an IBM quantum device. The experiment was carried out on the first 8 qubits of the device \textit{ibmq\_guadalupe}. This is a quantum device of 16 qubits, 32 of quantum volume, and $1.245\times 10^{-2}$ of average error in the CNOT gates. The VQE was implemented with the same configuration as the simulations, that is, the same Hamiltonian, ansatz, and classical optimizer. Figure~\ref{fig:VQE_result}~b) shows the results of the experiments. We can see that TPB and HEEM outperform EM in precision at the same number of iterations, while HEEM provides a slight advantage over TPB. The realization of TPB was very favorable, being approximately one standard deviation below the mean shown in the simulations. Although HEEM performs similarly to TPB in the number of iterations, the scenario differs in the number of circuits. The inset of Fig.~\ref{fig:VQE_result}~b), which are results of experimental VQEs in terms of the number of circuits, shows that HEEM overcomes both TPB and EM in the number of circuits, achieving better energies at the same number of circuits. Grouping with HEEM also provides an advantage over TPB and EM at runtime. The time for each experiment was 3.8 hours with TPB, 4.4 hours with EM, and 2.7 hours with HEEM. Thus, HEEM provides a speed-up with respect to the other grouping approaches in real hardware.  
	
	\section{Conclusions and Outlook}
    In the NISQ era, in order to achieve a quantum advantage with variational algorithms, we need efficient techniques to evaluate the expected values of Hamiltonians. In this article, we introduce the Hardware Efficient Entangled Measurements (HEEMs). They allow one to evaluate the expected value of a Hamiltonian by simultaneously measuring groups of Pauli strings and only employing entangled gates between qubits that are physically connected on the device. This makes HEEM an efficient and noise-robust alternative for evaluating expected values and speeding up variational algorithms in NISQ devices. Since there are multiple ways to group a set of Pauli strings with HEEM, we have introduced three algorithms to carry out the grouping. The first is the \emph{naive} HEEM, which does not optimize the order of the grouping loops. The other two are \emph{order-connected} HEEM and the \emph{order-disconnected} HEEM. They optimize the grouping loops by using the \textit{processor embedding} problem, the problem of finding the optimal map of the theoretical qubits into the physical qubits. 
	
	We compare our methods and conclude that the \emph{order-connected} and \emph{order-disconnected} algorithms outperform the \emph{naive} algorithm on the average number of groups and the time execution. Our methods improve previous works, using fewer groups than TPB \cite{bravyi2017}, with a slight increase in the number of required CNOTs. Furthermore, HEEM requires fewer CNOTs than EM \cite{hamamura2020efficient}, using only a few more groups. Note that the number of groups quantifies the statistical error, while the number of CNOTs quantifies the experimental error. Then, taking both sources of errors into account, HEEM outperforms both TPB and EM. We have shown, with noisy simulations and experiments, that HEEM achieves better results with the same number of circuits. In addition, HEEM is faster than TPB and EM to be implemented both in an experiment and in a simulation for practical scenarios.	HEEM grouping can be useful not only in variational algorithms, but also in any task that requires the evaluation of several Pauli strings, such as full quantum tomography \cite{qubit_tomography,Hffner2005}, compressed sensing \cite{Gross2010,Riofro2017}, reduced density matrix tomography \cite{Cotler2020, Bonet-Monroig2020}, classical shadows \cite{Huang2020,Chen2021}, or direct fidelity estimation \cite{DFE1,DFE2}.
	
	Several extensions can be incorporated into our proposal. The grouping of Pauli strings by HEEM can be improved by considering entanglement between a larger number of qubits. This would allow for further reduction of the number of groups at the expense of more entanglement resources. The method can also be refined by including the error of the CNOT gates of the chip as weights in the graph that represents the connectivity of the chip. This would produce better results by ensuring that the majority of entangling operations are performed over pairs of qubits with the lowest CNOT gate errors. Given that the number of groups and CNOTs are proxies for the error of the measurement, we can explore the behavior of HEEM using more elaborate metrics, such as those proposed in \cite{crawford2021efficient}.  In this article, it is suggested to sort the Pauli strings according to their weights in the Hamiltonian, which could improve HEEM in practical scenarios. The combination of HEEM with other estimation protocols could provide an even more accurate estimate of observables on current devices, such as error mitigation techniques \cite{mitigation1,mitigation2,mitigation3} or adaptive schemes \cite{2110.15339}. 

	The code for reproducing the algorithms and figures in this article can be found in \cite{github}.
	
	\section{Acknowledgements}
	
	The authors thank Juan José García-Ripoll for valuable feedback on the manuscript. This work has been supported by the CSIC Interdisciplinary Thematic Platform (PTI+) on Quantum Technologies (PTI-QTEP+). G.~F.~P. and G.~J. acknowledge support from the European Union's Horizon 2020 FET-Open project SuperQuLAN (899354). F.~E.~G. was supported by a Marie Sk{\l}odowska-Curie Action from the EC (COFUND grant no. 945045), and by the NWO Gravitation project NETWORKS (grant no. 024.002.003). D.~F.~F. acknowledges support from the FPU Program No. FPU20/04762. L.~P. was supported by ANID-PFCHA/DOCTORADO-BECAS-CHILE/2019-77220027, CAM/FEDER Project No. S2018/TCS-4342 (QUITEMAD-CM), and the Proyecto Sinergico CAM 2020 Y2020/TCS-6545 (NanoQuCo-CM). The authors thank the IBM Quantum Team for making multiple devices available to the CSIC-IBM Quantum Hub via the IBM Quantum Experience. The views expressed are those of the authors and do not reflect the official policy or position of IBM or the IBM Quantum team.
	
    \bibliography{Bibliography.bib}
	
    \onecolumngrid\newpage
    \appendix
	
    \begin{center}
        \textbf{\large Supplemental Material \\ ~ \\Hardware-efficient entangled measurements for variational quantum algorithms}
    \end{center}
	\begin{center}
		Francisco Escudero,${}^{1,2}$ David Fern{\'a}ndez-Fern{\'a}ndez,${}^{1,3}$, Gabriel Jaum{\`a},${}^1$ Guillermo F. Pe{\~n}as,${}^1$ Luciano Pereira${}^1$\\
		\vspace{0.2cm}{\small $^1${\em Instituto de F{\'i}sica Fundamental, IFF-CSIC, Calle Serrano 113b, 28006 Madrid, Spain}\\
			${}^2${\em Networks, Qusoft and CWI, Amsterdam, Netherlands.}\\
			${}^3${\em Instituto de Ciencia de Materiales de Madrid, ICMM-CSIC, 28049 Madrid, Spain}}
	\end{center}

    \setcounter{figure}{0}
    \setcounter{table}{0}

	\section{Entangled measurements}\label{sec:ApenMeasurments}
	
	We have identified the two-qubit measurements that allow to measure all maximal sets of compatible Pauli strings of length two. They are the following ones:
	
	\vspace{0.5cm}
	
	\begin{quantikz}
		\text{Bell measurement} & & &\\
		\lstick{$q_0$ :} & \ctrl{1} & \gate{H}  &  \meter{} \\
		\lstick{$q_1$ :  } & \targ{}  & \qw & \meter{}
	\end{quantikz} \hspace{2.5cm}
	\begin{quantikz}
		 \text{$\Omega^{X}$ measurement}& & & & & \\
		\lstick{$q_0$ :  } & \gate{S} & \gate{H} & \ctrl{1} & \gate{H}  &  \meter{} \\
		\lstick{$q_1$ :  } & \gate{S} & \qw      &  \targ{} & \qw      &  \meter{}
	\end{quantikz}
	\vspace{.5cm}
	
	\begin{quantikz}
		 \text{$\Omega^{Y}$ measurement}& & &\\
		\lstick{$q_0$ :  }  & \gate{H} & \ctrl{1} & \gate{H}  &  \meter{} \\
		\lstick{$q_1$ :  }  & \qw      &  \targ{} & \qw      &  \meter{}
	\end{quantikz}\hspace{1.3cm}
	\begin{quantikz}
		 \text{$\Omega^{Z}$ measurement} & & & \\
		\lstick{$q_0$ :  }  & \gate{S} & \ctrl{1} & \gate{H}  &  \meter{} \\
		\lstick{$q_1$ :  }  & \qw      &  \targ{} & \qw       &  \meter{}
	\end{quantikz}
	\vspace{.5cm}
	
	\begin{quantikz}
		\text{$\chi$ measurement}&  & &\\
		\lstick{$q_0$ :  }  & \gate{U_2(\pi/2, \pi)} & \ctrl{1} & \gate{H}  &  \meter{} \\
		\lstick{$q_1$ :  }  & \qw                    & \targ{} & \qw        & \meter{}
	\end{quantikz} \hspace{.5cm}
	\begin{quantikz}
		\text{$\tilde{\chi}$ measurement}& & &\\
		\lstick{$q_0$ :  }  & \gate{U_2(0, \pi/2)} & \ctrl{1} & \gate{H}  &  \meter{} \\
		\lstick{$q_1$ :  }  & \qw                  &  \targ{} & \qw       &  \meter{}
	\end{quantikz}
	
	\vspace{1cm}

 In the last two entangled measurements, the gate $U_2$ refers to the following operation in matrix form 
 $$
 U_2(\phi, \lambda) = \frac{1}{\sqrt{2}}
    \begin{pmatrix}
        1          & -e^{i\lambda} \\
        e^{i\phi} & e^{i(\phi+\lambda)}
    \end{pmatrix}.$$
 
	\subsection{Computing the expected values of Hamiltonians with a certain grouping}\label{computing_exp_val}
	
	In this section, we explain how to obtain the expected value of a Hamiltonian once we have grouped its Pauli strings. Let $H=\sum_\alpha h_\alpha P_\alpha$ be a multi-qubit Hamiltonian and suppose that the $m$ first Pauli strings are compatible with a single HEEM measurement into the basis $\mathcal{B}=\mathcal{B}_1\otimes\dots\otimes \mathcal{B}_t$, with $B_i$ single- or two-qubit bases with $i=1,\dots,t$. This means that the Pauli strings $P_\alpha$ with $\alpha=1,\dots,m$ are diagonal on the basis of $\mathcal{B}$. Let be $\vec{W}_{P} = {\rm Diag}_{\mathcal{B}}\{P\}$ the vector with the diagonal entries of the Pauli string $P$ on the basis of $\mathcal{B}$. The elements of $\vec{W}_{P}$ correspond to the eigenvalues of $P$, each of them repeated by its multiplicity. Therefore, if $\vec{\mathcal{P}}$ is the probability distribution of a measurement on the basis $\mathcal{B}$, the expected value of the compatible Pauli strings is given by
	\begin{equation}\label{expected_value}
	    \left\langle\sum_{\alpha=1}^m h_\alpha P_\alpha\right\rangle =\left( \sum_{\alpha=1}^m h_\alpha \vec{W}_{P_\alpha},\vec{\mathcal{P}}\right),
	\end{equation}
	where $(\cdot,\cdot)$ is the usual inner product. To illustrate the procedure, we evaluate the energy of a simple Hamiltonian consisting only of two Pauli strings,
	 \begin{equation}\label{eq:toyHamiltonian}
	     	     H = 2IZY + 4ZXZ.
	 \end{equation}
	We have to check in Table~\ref{table1} on which basis, if possible, these two strings can be measured together. After a careful look, we realize that the operators $I$ and $Z$ of the first qubits are compatible with $\mathcal{Z}$, and that the operators $ZY$ and $XZ$ of the second and third qubits are compatible with $\tilde{\chi}$. The basis $\mathcal{Z}$ allows us to diagonalize the operators $ I  = +\ket{0}\bra{0} + \ket{1} \bra{1},$ and $ Z  = +\ket{0}\bra{0} - \ket{1} \bra{1}$, while the basis $\tilde{\chi}$ the 2-qubit operators 
    \begin{align}\label{eq:expected}
    YX & = -\ket{\tilde{\chi}_0}\bra{\tilde{\chi}_0} +\ket{\tilde{\chi}_1}\bra{\tilde{\chi}_1} +\ket{\tilde{\chi}_2}\bra{\tilde{\chi}_2} -\ket{\tilde{\chi}_3}\bra{\tilde{\chi}_3}, \\
    ZY & = +\ket{\tilde{\chi}_0}\bra{\tilde{\chi}_0} +\ket{\tilde{\chi}_1}\bra{\tilde{\chi}_1} -\ket{\tilde{\chi}_2}\bra{\tilde{\chi}_2} -\ket{\tilde{\chi}_3}\bra{\tilde{\chi}_3}, \\
    XZ  & = +\ket{\tilde{\chi}_0}\bra{\tilde{\chi}_0} -\ket{\tilde{\chi}_1}\bra{\tilde{\chi}_1} +\ket{\tilde{\chi}_2}\bra{\tilde{\chi}_2} -\ket{\tilde{\chi}_3}\bra{\tilde{\chi}_3}.
   \end{align}
    One last important thing that we need is the explicit expression of the vectors of the $\tilde{\chi}$ basis in the computational basis, to correctly relate the vector of weights with the vector of outcomes
    \begin{align}
   |\tilde{\chi} _0\rangle &= +i|00\rangle-|01\rangle+i|10\rangle+|11\rangle, \\
   |\tilde{\chi} _1\rangle &= +|00\rangle+i|01\rangle-|10\rangle+i|11\rangle, \\
   |\tilde{\chi} _2\rangle &= +i|00\rangle+|01\rangle+i|10\rangle-|11\rangle, \\
   |\tilde{\chi} _3\rangle &= -|00\rangle+i|01\rangle+|10\rangle+i|11\rangle.
   \end{align} 
    See subsection~\ref{Sumary} for explicit expressions of how to construct every entangled measurement and the diagonal representation of pairs of observables in all of these bases. Now we can tackle the problem of evaluating the energy of Eq.~\eqref{eq:toyHamiltonian} with a measurement on the basis $\mathcal{B}=\mathcal{Z}_1\otimes\tilde{\chi}_{2,3}$, where the subscripts refer to the subspace spanned by such qubits. For the Pauli string $IZY$, we have $\text{Diag}_{\mathcal{Z}} \left\{ I \right\} = [+1,+1]$ and $\text{Diag}_{\tilde{\chi}} \left\{ ZY \right\} = [+1,+1,-1,-1]$. The vector $\Vec{W}_{IZY}$ is computed as the Kronecker product of the vectors $\text{Diag}_{\mathcal{Z}}$ and $\text{Diag}_{\tilde{\chi}}$. Thus,
   \begin{equation}
       h_{IZY}\Vec{W}_{IZY} = 2\times [+1,+1] \otimes [+1,+1,-1,-1] = [+2,+2,-2,-2,+2,+2,-2,-2].
   \end{equation}
   Analogously, for the string $ZXZ$, we have $\text{Diag}_{\mathcal{Z}} \left\{Z \right\} = [+1,-1]$ and $ \text{Diag}_{\tilde{\chi}} \left\{XZ \right\} = [+1,-1,+1,-1] $. Then
    \begin{equation}
       h_{ZXZ}\vec{W}_{ZXZ} = 4\times [+1,-1] \otimes [+1,-1,+1,-1] = [+4,-4,+4,-4,-4,+4,-4,+4].
   \end{equation}
   Therefore, we have that
   \begin{equation}
       \sum_{\alpha=1}^m h_\alpha \vec{W}_{P_\alpha} = [+6,-2,+2,-6,-2,+6,-6,+2].
   \end{equation}
   In summary, what our algorithm does once it has identified which Pauli strings can be measured together is to check with which measurement basis it can do so and in which order these basis vectors have to be taken into account. After sorting out all the plus and minus signs, all that is left to do is multiply by the weight in the Hamiltonian $h_{\alpha}$, and finally plug in the actual result of the experiment with Eq.~\eqref{expected_value}. 

   \begin{table}[h!]
	\centering
		\resizebox{0.4\columnwidth}{!}{
			\begin{tabular}{c|cccccccc|c|}
				\cline{2-10}
				& \multicolumn{1}{c|}{$XX$} & \multicolumn{1}{c|}{$YZ$} & \multicolumn{1}{c|}{$ZY$} & \multicolumn{1}{c|}{$YY$} & \multicolumn{1}{c|}{$XZ$} & \multicolumn{1}{c|}{$ZX$} & \multicolumn{1}{c|}{$ZZ$} & $XY$ & $YX$ \\ \hline
				\multicolumn{1}{|c|}{$XX$} & \multicolumn{1}{c|}{---}  & \multicolumn{1}{c|}{$\Omega^X$} & \multicolumn{1}{c|}{$\Omega^X$} & \multicolumn{1}{c|}{Bell} & \multicolumn{1}{c|}{\xmark} & \multicolumn{1}{c|}{\xmark} & \multicolumn{1}{c|}{Bell} & \xmark & \xmark \\ \hline
				\multicolumn{1}{|c|}{$YZ$} & \multicolumn{1}{c|}{} & \multicolumn{1}{c|}{---}& \multicolumn{1}{c|}{$\Omega^X$} & \multicolumn{1}{c|}{\xmark} & \multicolumn{1}{c|}{\xmark} & \multicolumn{1}{c|}{$\chi$} & \multicolumn{1}{c|}{\xmark} & $\chi$ & \xmark \\ \cline{1-1} \cline{3-10} 
				\multicolumn{1}{|c|}{$ZY$} & & \multicolumn{1}{c|}{}   & \multicolumn{1}{c|}{---}& \multicolumn{1}{c|}{\xmark} & \multicolumn{1}{c|}{$\tilde{\chi} $} & \multicolumn{1}{c|}{\xmark} & \multicolumn{1}{c|}{\xmark} & \xmark & $\tilde{\chi} $ \\ \cline{1-1} \cline{4-10} 
				\multicolumn{1}{|c|}{$YY$} & & & \multicolumn{1}{c|}{}   & \multicolumn{1}{c|}{---}& \multicolumn{1}{c|}{$\Omega^Y$} & \multicolumn{1}{c|}{$\Omega^Y$} & \multicolumn{1}{c|}{Bell} & \xmark & \xmark \\ \cline{1-1} \cline{5-10} 
				\multicolumn{1}{|c|}{$XZ$} & & & & \multicolumn{1}{c|}{}   & \multicolumn{1}{c|}{---}& \multicolumn{1}{c|}{$\Omega^Y$} & \multicolumn{1}{c|}{\xmark} & \xmark & $\tilde{\chi} $ \\ \cline{1-1} \cline{6-10} 
				\multicolumn{1}{|c|}{$ZX$} & & & & & \multicolumn{1}{c|}{}   & \multicolumn{1}{c|}{---}& \multicolumn{1}{c|}{\xmark} & $\chi$ & \xmark \\ \cline{1-1} \cline{7-10} 
				\multicolumn{1}{|c|}{$ZZ$} & & & & & & \multicolumn{1}{c|}{}   & \multicolumn{1}{c|}{---}& $\Omega^Z$ & $\Omega^Z$ \\ \cline{1-1} \cline{8-10} 
				\multicolumn{1}{|c|}{$XY$} & & & & & & & \multicolumn{1}{c|}{}   & ---& $\Omega^Z$ \\ \cline{1-1} \cline{9-10} 
				\multicolumn{1}{|c|}{$YX$} & & & & & & & & & ---\\ \cline{1-1} \cline{10-10} \hline  
		\end{tabular}}
		\caption{Compatibility relation between two-qubit Pauli strings. The symbol \xmark\ indicates that the corresponding Pauli strings are not jointly measurable. The other boxes contain the entangled measurements that are compatible with the corresponding Pauli strings. See Appendix~\ref{Sumary} for the definitions of these measurements.}
		\label{table1}
	\end{table}
   
\subsection{Jointly diagonalizable pairs in all entangled bases.}\label{Sumary}

\begin{align*}
	\mathrm{Bell} :  &\\ 
|\Phi_0\rangle = |00\rangle + |11\rangle & \hspace{1.3cm} \text{Commuting pairs} \\ 
|\Phi_1\rangle = |00\rangle-|11\rangle & \hspace{1.5cm}    XX  = +\ket{\Phi_0}\bra{\Phi_0} -\ket{\Phi_1}\bra{\Phi_1} +\ket{\Phi_2}\bra{\Phi_2}-\ket{\Phi_3}\bra{\Phi_3} &\\
   |\Phi_2\rangle = |01\rangle + |10\rangle & \hspace{1.5cm}  YY   = -\ket{\Phi_0}\bra{\Phi_0} +\ket{\Phi_1}\bra{\Phi_1} +\ket{\Phi_2}\bra{\Phi_2}-\ket{\Phi_3}\bra{\Phi_3} &\\
   |\Phi_3\rangle = |01\rangle-|10\rangle & \hspace{1.5cm}   ZZ   = +\ket{\phi_0}\bra{\Phi_0} +\ket{\Phi_1}\bra{\Phi_1} -\ket{\Phi_2}\bra{\Phi_2}-\ket{\Phi_3}\bra{\Phi_3}  &\\
   &\\
	\Omega^X :  &\\ 
   |\Omega^{X}_0\rangle= +|00\rangle-i|01\rangle-i|10\rangle+|11\rangle & \hspace{1.3cm} \text{Commuting pairs} \\
   |\Omega^{X}_1\rangle= +|00\rangle-i|01\rangle+i|10\rangle-|11\rangle & \hspace{1.5cm}  
    YZ  = -\ket{\Omega^{X}_0}\bra{\Omega^{X}_0} +\ket{\Omega^{Y}_1}\bra{\Omega^{X}_1} -\ket{\Omega^{X}_2}\bra{\Omega^{X}_2}+\ket{\Omega^{X}_3}\bra{\Omega^{X}_3} &\\
   |\Omega^{X}_2\rangle= +|00\rangle+i|01\rangle-i|10\rangle-|11\rangle & \hspace{1.5cm} 
   XX   = +\ket{\Omega^{X}_0}\bra{\Omega^{X}_0} -\ket{\Omega^{X}_1}\bra{\Omega^{X}_1} -\ket{\Omega^{X}_2}\bra{\Omega^{X}_2} + \ket{\Omega^{X}_3}\bra{\Omega^{X}_3} &\\
   |\Omega^{X}_3\rangle= -|00\rangle-i|01\rangle-i|10\rangle-|11\rangle & \hspace{1.5cm}  
    ZY   = -\ket{\Omega^{X}_0}\bra{\Omega^{X}_0} -\ket{\Omega^{X}_1}\bra{\Omega^{X}_1} +\ket{\Omega^{X}_2}\bra{\Omega^{X}_2}+\ket{\Omega^{X}_3}\bra{\Omega^{X}_3} & \\
   &\\
	\Omega^Y: &\\
   |\Omega^{Y}_0\rangle= +|00\rangle+|01\rangle+|10\rangle-|11\rangle   & \hspace{1.3cm} \text{Commuting pairs}\\
   |\Omega^{Y}_1\rangle= +|00\rangle+|01\rangle-|10\rangle+|11\rangle & \hspace{1.2cm}  
     XZ  = +\ket{\Omega^{Y}_0}\bra{\Omega^{Y}_0} -\ket{\Omega^{Y}_1}\bra{\Omega^{Y}_1} +\ket{\Omega^{Y}_2}\bra{\Omega^{Y}_2} - \ket{\Omega^{Y}_3}\bra{\Omega^{Y}_3} &\\
   |\Omega^{Y}_2\rangle= +|00\rangle-|01\rangle+|10\rangle+|11\rangle & \hspace{1.2cm}  
    YY  = +\ket{\Omega^{Y}_0}\bra{\Omega^{Y}_0} -\ket{\Omega^{Y}_1}\bra{\Omega^{Y}_1} -\ket{\Omega^{Y}_2}\bra{\Omega^{Y}_2}+\ket{\Omega^{Y}_3}\bra{\Omega^{Y}_3} &  \\
   |\Omega^{Y}_3\rangle= -|00\rangle+|01\rangle+|10\rangle+|11\rangle & \hspace{1.2cm}  
     ZX  = +\ket{\Omega^{Y}_0}\bra{\Omega^{Y}_0} +\ket{\Omega^{Y}_1}\bra{\Omega^{Y}_1} -\ket{\Omega^{Y}_2}\bra{\Omega^{Y}_2}-\ket{\Omega^{Y}_3}\bra{\Omega^{Y}_3} & \\
    &\\
     \Omega^Z: & \\
     |\Omega^{Z}_{0}\rangle= +|00\rangle - i|11\rangle &  \hspace{1.3cm}\text{Commuting pairs} \\
   |\Omega^{Z}_{1}\rangle= +|00\rangle + i|11\rangle & \hspace{1.2cm}  
     XY  = -\ket{\Omega^{Z}_0}\bra{\Omega^{Z}_0} +\ket{\Omega^{Z}_1}\bra{\Omega^{Z}_1} -\ket{\Omega^{Z}_2}\bra{\Omega^{Z}_2} + \ket{\Omega^{Z}_3}\bra{\Omega^{Z}_3} & \\
   |\Omega^{Z}_{2}\rangle= +|01\rangle + i|10\rangle & \hspace{1.2cm}  
     YX  = -\ket{\Omega^{Z}_0}\bra{\Omega^{Z}_0} +\ket{\Omega^{Z}_1}\bra{\Omega^{Z}_1} +\ket{\Omega^{Z}_2}\bra{\Omega^{Z}_2} - \ket{\Omega^{Z}_3}\bra{\Omega^{Z}_3} & \\
   |\Omega^{Z}_{3}\rangle= +|00\rangle - i|10\rangle & \hspace{1.2cm}  
     ZZ  = +\ket{\Omega^{Z}_0}\bra{\Omega^{Z}_0} +\ket{\Omega^{Z}_1}\bra{\Omega^{Z}_1} -\ket{\Omega^{Z}_2}\bra{\Omega^{Z}_2} - \ket{\Omega^{Z}_3}\bra{\Omega^{Z}_3} & \\
    &\\
     \chi:  \\
     |\chi_0\rangle= -|00\rangle+|01\rangle-i|10\rangle+i|11\rangle   &  \hspace{1.3cm}\text{Commuting pairs} \\
   |\chi_1\rangle= +|00\rangle+|01\rangle+i|10\rangle-i|11\rangle     &     \hspace{1.2cm}  
     XY  = +\ket{\chi_0}\bra{\chi_0} -\ket{\chi_1}\bra{\chi_1} +\ket{\chi_2}\bra{\chi_2} - \ket{\chi_3}\bra{\chi_3} & \\
   |\chi_2\rangle= +|00\rangle-|01\rangle+i|10\rangle+i|11\rangle     &     \hspace{1.2cm}  
     YZ  = -\ket{\chi_0}\bra{\chi_0} +\ket{\chi_1}\bra{\chi_1} +\ket{\chi_2}\bra{\chi_2} - \ket{\chi_3}\bra{\chi_3} & \\
   |\chi_3\rangle= -|00\rangle+|01\rangle+i|10\rangle+i|11\rangle     &     \hspace{1.2cm}  
     ZX  = +\ket{\chi_0}\bra{\chi_0} +\ket{\chi_1}\bra{\chi_1} -\ket{\chi_2}\bra{\chi_2} - \ket{\chi_3}\bra{\chi_3} &\\
     &\\
    \tilde\chi:  \\
     |\tilde\chi_0\rangle= +i|00\rangle-|01\rangle+i|10\rangle+|11\rangle   &  \hspace{1.3cm}\text{Commuting pairs} \\
   |\tilde\chi_1\rangle= +|00\rangle+i|01\rangle-|10\rangle+i|11\rangle     &     \hspace{1.2cm}  
     YX  = -\ket{\tilde\chi_0}\bra{\tilde\chi_0} + \ket{\tilde\chi_1}\bra{\tilde\chi_1} +\ket{\tilde\chi_2}\bra{\tilde\chi_2} - \ket{\tilde\chi_3}\bra{\tilde\chi_3} & \\
   |\tilde\chi_2\rangle= +i|00\rangle+|01\rangle+i|10\rangle-|11\rangle     &     \hspace{1.2cm}  
     ZY  = +\ket{\tilde\chi_0}\bra{\tilde\chi_0} +\ket{\tilde\chi_1}\bra{\tilde\chi_1} -\ket{\tilde\chi_2}\bra{\tilde\chi_2} - \ket{\tilde\chi_3}\bra{\tilde\chi_3} & \\
   |\tilde\chi_3\rangle= -|00\rangle+i|01\rangle+|10\rangle+i|11\rangle     &     \hspace{1.2cm}  
     XZ  = +\ket{\tilde\chi_0}\bra{\tilde\chi_0} - \ket{\tilde\chi_1}\bra{\tilde\chi_1} + \ket{\tilde\chi_2}\bra{\tilde\chi_2} - \ket{\tilde\chi_3}\bra{\tilde\chi_3} &	
   \end{align*}
	
	\clearpage
	
	\section{Algorithms}\label{algGrouping}
	In this appendix, we present the heuristic algorithms announced in the main text. All of them can be found in \cite{github}. For the pseudo-code shown in this appendix, we only use variables representing integers, booleans, strings, and lists. Other data structures, such as hash maps or queues, can be used to improve performance. However, we stick to the most common data structures to increase code readability. Items within the lists can be integers or other lists to represent matrices. To index the lists, we use square brackets as $List[i]$, with the indices starting at 0. Lists are created as $List = [item1, item2, \ldots]$. When calling other functions, we use parentheses as FunctionName(parameter1, parameter2, $\ldots$). Finally, comments inside the codes start with //. Note that the notation used in this appendix may be different from the one shown in the main text, this is so as to have a notation closer to common programming languages in the pseudo codes, thus facilitating their understanding.
    
    \subsection{Grouping algorithms}
\begin{figure}[ht!]
		\centering
		\includegraphics[width=0.8\linewidth]{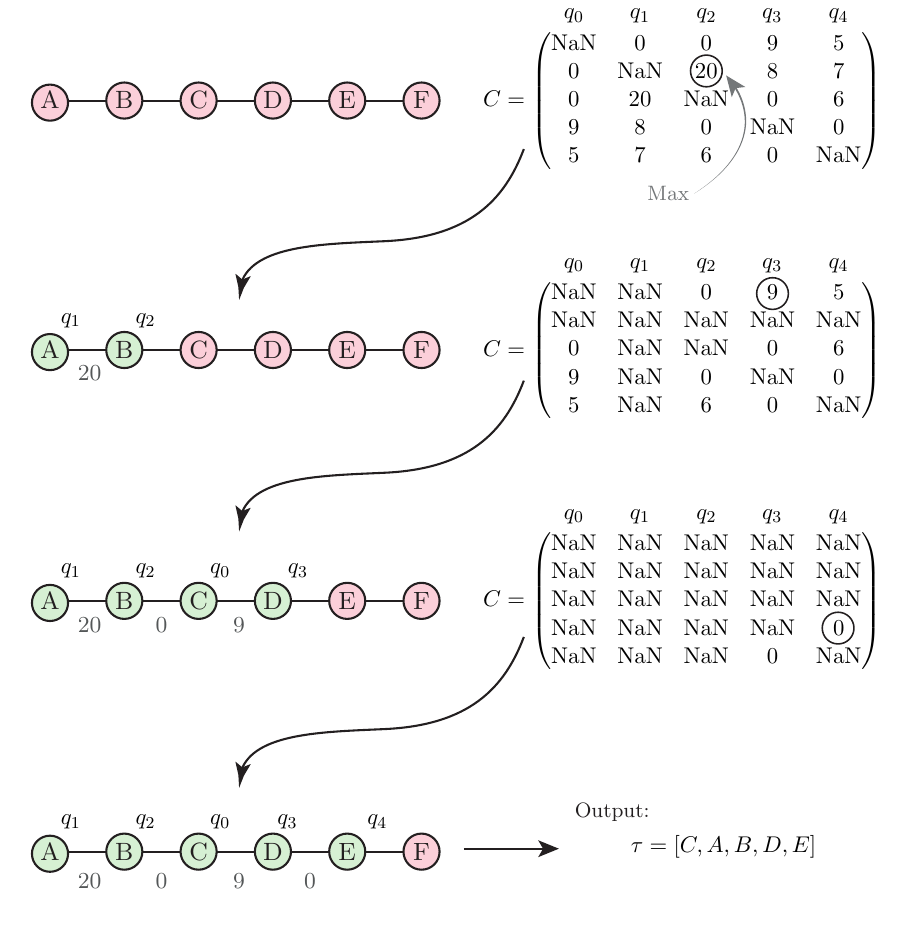}
		\caption{\textbf{Example of Subroutine~\ref{sub:3theophysNoncon} (Processor embedding map construction, without ensuring that $\tau(G')$ is connected).} The subroutine begins by mapping the pair of theoretical qubits $(q_1,q_2)$, corresponding to the highest entry of the compatibility matrix $C$, to a pair of physically connected qubits. Then, it updates the compatibility matrix, setting to NaN all the entries corresponding to $q_1$, as it has been mapped to a physical qubit with no more available connections. After that, it repeats this step two more times and determines a $\tau$ with a total number of $20+0+9+0=29$ compatibilities, which is 11 less than what the Subroutine~\ref{sub:4theophysConnected} would get, as shown in Figure~\ref{fig:subroutine_4}. Note that the physical connections B-C and D-E are not going to be useful to measure entangled pairs, which is something Subroutine~\ref{sub:4theophysConnected} precludes.}
		\label{fig:subroutine_3}
	\end{figure}

\begin{figure}[ht!]
		\centering
		\includegraphics[width=0.8\linewidth]{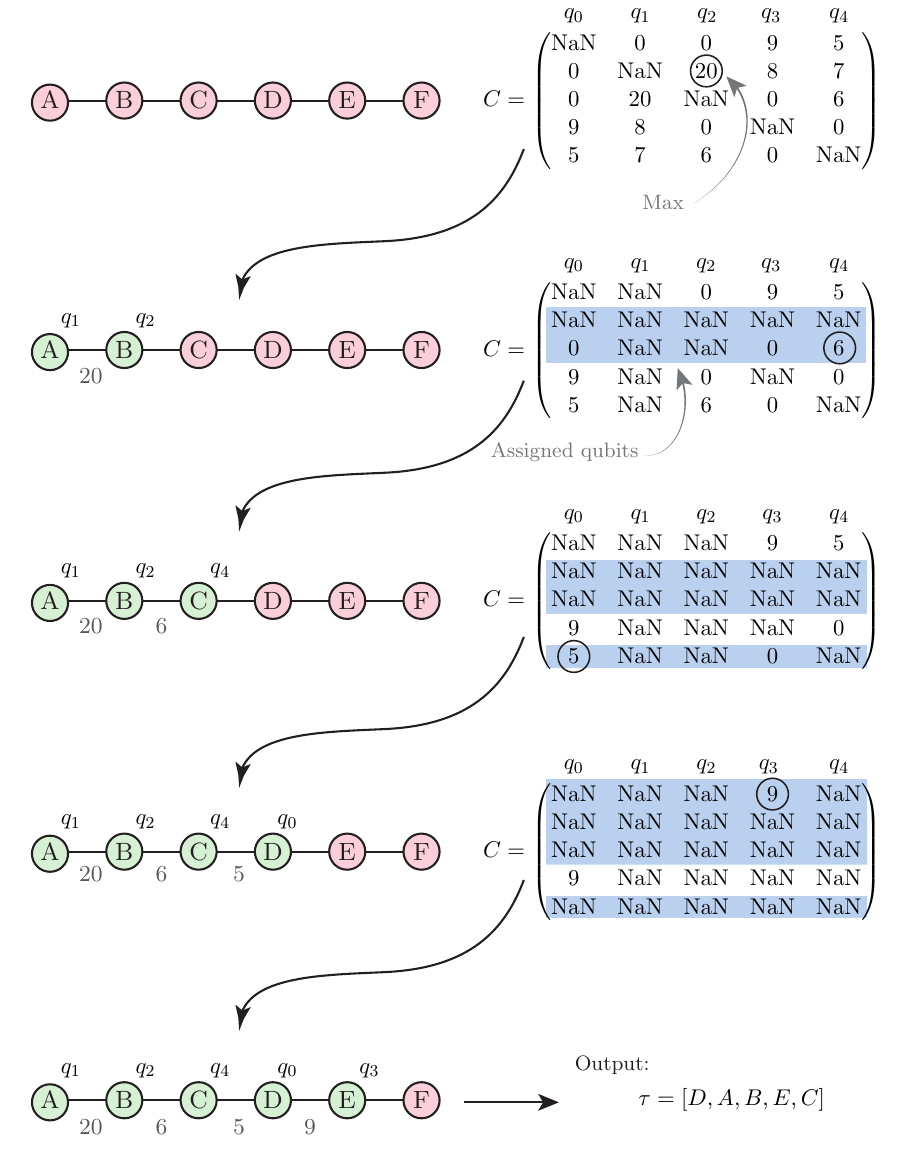}
		\caption{\textbf{Example of Subroutine~\ref{sub:4theophysConnected} (Processor embedding map construction, ensuring that $\tau(G')$ is connected).} The subroutine begins by mapping the pair of theoretical qubits $(q_1,q_2)$, corresponding to the highest entry of the compatibility matrix $C$, to a pair of physically connected qubits. Then it updates the compatibility matrix, setting NaN for all entries corresponding to $q_1$, since it has been assigned to a physical qubit with no more available connections. Then, we search for the entry with maximum compatibility among the ones shared with the assigned qubits (blue shadow). This step ensures that no physical connection in the chip is wasted. The same procedure is repeated until all theoretical qubits have been assigned and ends up with a $\tau$ that has $20+6+5+9=40$ compatibilities, 11 more than Subroutine~\ref{sub:3theophysNoncon}, as shown in Figure~\ref{fig:subroutine_3}.}
		\label{fig:subroutine_4}
	\end{figure}

    All our grouping algorithms (\emph{naive}, \emph{order-connected}, and \emph{order-disconnected}) take into account the connectivity of the chip, but differ in the way theoretical qubits are mapped onto physical qubits. The \emph{naive} algorithm (Algorithm~\ref{alg:NaiveGrouping'}) maps the $i$-th theoretical qubit to the $i$-th physical qubit (i.e., it chooses the trivial \textit{processor embedding} map $\tau[i]=i$). The \emph{order-disconnected} (Algorithm~\ref{Alg:OrdDisc'}) uses a subroutine that generates a \textit{processor embedding} map $\tau$ that maps the theoretical qubits to the physical ones seeking a high number of measurement compatibilities. The \emph{order-connected} (Algorithm~\ref{Alg:OrdConnected'}) does the same as the \emph{order-disconnected}, but with the additional requirement that the graph $G'$ of theoretical qubits is connected once it is mapped to the graph $G$ of physical qubits through $\tau$, i.e., $\tau
    [G']$ is a connected subgraph of $G$ (which can always be done if $G$ is connected and has at least the same number of vertices as $G'$). This requirement aims to ensure that all connections of the chip are useful to measure, as explained in Figures~\ref{fig:subroutine_3} and \ref{fig:subroutine_4}.

    The core of our algorithms is Subroutine~\ref{sub:1NaiveGrouping}, which takes the following inputs: a set of Pauli strings, a \textit{processor embedding} map $\tau$, and the order in which the subroutine runs over the qubits and measurements. All algorithms use Subroutine~\ref{sub:6tauCompMat} to decide the orders in which Subroutine~\ref{sub:1NaiveGrouping} runs over the qubits and measurements. The \emph{naive} algorithm does that assuming the trivial \textit{processor embedding} map $\tau[i]=i$, while the \emph{order-disconnected} algorithm uses Subroutine~\ref{sub:3theophysNoncon} to design $\tau$ and the \emph{order-connected} uses Subroutine~\ref{sub:4theophysConnected} for the same purpose. 
    
    Finally, we would like to remark that despite that Algorithm~\ref{Alg:OrdConnected'} may look like an arbitrary modification of Algorithm~\ref{Alg:OrdDisc'}, the results show that performance improves when imposing that $\tau[G']$ is connected (see Figure~\ref{fig:violinHEEM} of the main text or Figures~\ref{fig:subroutine_3} and \ref{fig:subroutine_4}). 

    \begin{algorithm}[h!]\label{alg:NaiveGrouping'}
		\caption{Naive grouping}
		\KwIn{$n$ Pauli strings $PS$ of $N$ qubits and chip's connectivity $G$}
		
		Define the trivial map $\tau[i]=i$
  
		$CM^\tau$, $CQ^\tau$ = Orders($PS$, $\tau$, $G$) \space \textcolor{blue}{// (Subroutine~\ref{sub:6tauCompMat}) $CM^\tau$ encodes the measurements and $CQ^\tau$ the qubits}
		
		$M$, $Gr$ = Grouping($PS$, $\tau$, $G$,  $CM^\tau$, $CQ^\tau$) \space \textcolor{blue}{// (Subroutine~\ref{sub:1NaiveGrouping}) $M$ encodes the measurements and $Gr$ the groups}
  
		\Return{$M$, $Gr$} 
	\end{algorithm}

     \begin{algorithm}[h!]\label{Alg:OrdDisc'}
		\caption{order-disconnected grouping}
		\KwIn{$n$ Pauli strings $PS$ of $N$ qubits and chip's connectivity $G$}
		
		$\tau$ = DisonnectedMap($PS$, $G$) \space \textcolor{blue}{// Subroutine~\ref{sub:3theophysNoncon}}
  
		$CM^\tau$, $CQ^\tau$ = Order($PS$, $\tau$, $G$) \space \textcolor{blue}{// (Subroutine~\ref{sub:6tauCompMat}) $CM^\tau$ encodes the measurements and $CQ^\tau$ the qubits}
		
		$M$, $Gr$ = Grouping($PS$, $\tau$, $G$, $CM^\tau$, $CQ^\tau$) \space \textcolor{blue}{// (Subroutine~\ref{sub:1NaiveGrouping}) $M$ encodes the measurements and $Gr$ the groups}
  
		\Return{$M$, $Gr$} 
	\end{algorithm}

    \begin{algorithm}[h!]\label{Alg:OrdConnected'}
		\caption{order-connected grouping}
		\KwIn{$n$ Pauli strings $PS$ of $N$ qubits and chip's connectivity $G$}
		
		$\tau$ = ConnectedMap($PS$, $G$) \space \textcolor{blue}{// Subroutine~\ref{sub:4theophysConnected}}
  
		$CM^\tau$, $CQ^\tau$ = Orders($PS$, $\tau$, $G$) \space \textcolor{blue}{// (Subroutine~\ref{sub:6tauCompMat}) $CM^\tau$ encodes the measurements and $CQ^\tau$ the qubits}
		
		$M$, $Gr$ = Grouping($PS$, $\tau$, $G$,  $CM^\tau$, $CQ^\tau$) \space \textcolor{blue}{// (Subroutine~\ref{sub:1NaiveGrouping}) $M$ encodes the measurements and $Gr$ the groups}
  
		\Return{$M$, $Gr$} 
	\end{algorithm}

	\subsection{Subroutines}
    In this subsection, we present the subroutines used in Algorithms~\ref{alg:NaiveGrouping'}, \ref{Alg:OrdDisc'} and \ref{Alg:OrdConnected'}, and provide examples that illustrate how they work.
	\subsubsection{Grouping taking chip's connectivity into account}\label{sec:ApendConnecitivyAlgorithms}
	Herewith we describe the core subroutine of our grouping algorithms, which is inspired by the algorithm of \cite{hamamura2020efficient}. Our subroutine allows for a non-trivial \textit{processor embedding} map and optimizes the orders in which the algorithm runs through qubits and measurements, while the algorithm in \cite{hamamura2020efficient} does not.
	
	\begin{subroutine}[H]
		\caption{Greedy grouping taking chip's connectivity into account} \label{sub:1NaiveGrouping}
       \KwIn{$n$ Pauli strings $PS$, chip's connectivity $G$, \textit{processor embedding} map $\tau$, and vectors $CM^\tau$ and $CQ^\tau$ encoding the orders of measurements and qubits, respectively}
		
		Build the Pauli graph $PG$ of $PS$ \space \textcolor{blue}{// The elements $PG[i]$ are the Pauli strings}

        Sort $PG$ in descending order according to the nodes degree
        
		$M$ = [ ] \space \textcolor{blue}{// Measurements}
		
		$Gr$ = [ ] \space \textcolor{blue}{// Groups}
  
		\For{$i$ in [$0$, $\dots$, $n - 1$]\label{line:PauOrd1}}{
        
			\If{$i$ not in $Gr$}{
    
				$m$ = [ ]
				
				$gr$ = [$i$]
				
				\For{$j$ in [$i+1$, $\dots$, $n - 1$]\label{line:PauOrd2}}{
					\If{$j$ not in $Gr$}{
                        success, $\tilde m$ = CheckCompatibility($PG[i]$, $PG[j]$, $G$, $m$, $\tau$, $CM^\tau$, $CQ^\tau$) \space \textcolor{blue}{// Subroutine~\ref{sub:2assignMeas}}
                        
						\If{success}{
                            $m$ = $\tilde{m}$
                            
                            Append $j$ to $gr$
						}
					}
				}

                Let $U$ be the list of qubits not measured in $m$

                \For{$q$ in $U$}{

                Let $b$ be the basis compatible with the Pauli operator at position $q$ in $PG[i]$

                Append [$o$, $q$] to $m$
                }
				
				Append $m$ to $M$
    
                Append $gr$ to $Gr$
			}
		}
		\Return{$M$, $Gr$} 
	\end{subroutine}
 
    \vfill
    \newpage
    \begin{figure}[H]
		\centering
		\includegraphics[width=0.7\linewidth]{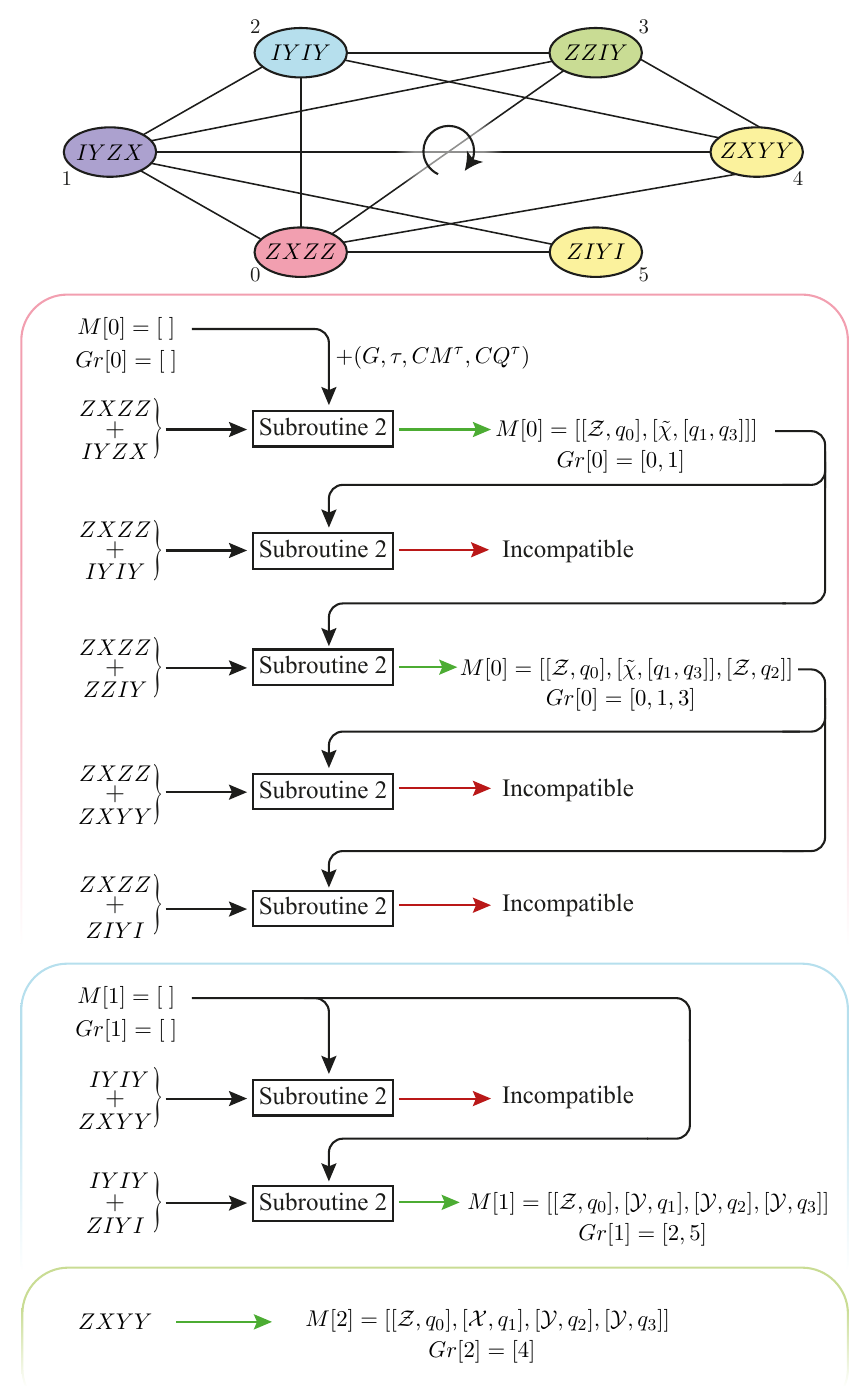}
		\caption{\textbf{Example of Subroutine~\ref{sub:1NaiveGrouping} (greedy Grouping taking chip's connectivity into account).} In the example, there are six Pauli strings to be grouped. On the top, the Pauli graph is depicted and on the bottom, a diagram details the algorithm used by Subroutine~\ref{sub:1NaiveGrouping}. On the one hand, if we used the LDFC algorithm to group the Pauli strings in the TPB, we would obtain five groups, represented by the five different colors in the Pauli graph. On the other hand, if we use Subroutine~\ref{sub:1NaiveGrouping} we obtain only three groups. The algorithm begins by sorting the Pauli strings in descending order of degree with respect to the Pauli graph. Then, it picks the highest degree string, $ZXZZ$ in our example, and tries to group it with the other strings using the Subroutine~\ref{sub:2assignMeas}. Note that this subroutine also needs the chip's connectivity $G$, the \textit{processor embedding} map $\tau$, the order for the measurements $CM^\tau$, and the order for the qubits $CQ^\tau$. However, these extra inputs are not depicted for simplicity. If the grouping is successful (green arrow), the Subroutine~\ref{sub:2assignMeas} outputs the measurements and the qubits to which it must be performed, encoded in $M[i]$, and updates the list that encodes the Pauli strings grouped $Gr[i]$. For the next iteration, the current measurements are provided to the Subroutine~\ref{sub:2assignMeas} so that all the grouped strings are compatible with each other. If grouping is not possible (red arrow), the algorithm does not update the measurements and passes to the next string. Once all Pauli strings are tested, we pick the next string with the highest degree which has not been grouped yet, and repeat the process until all strings are measured.}
		\label{fig:subroutine_1}
	\end{figure}

    \begin{subroutine}[H]
		\caption{Greedy measurement assignment taking chip's connectivity into account}\label{sub:2assignMeas}
		\KwIn{Two Pauli strings $PSi$ and $PSj$, chip's connectivity $G$, current assignment of measurements $m$, \textit{processor embedding} map $\tau$, and vectors $CM^\tau$ and $CQ^\tau$ encoding the orders of measurements and qubits, respectively}
        \If{ $PSj$ is not compatible with $m$}{
			\Return{\textbf{False}, $m$}}
   
        Copy $m$ in $\tilde m$  \space \textcolor{blue}{// Partial measurements}
        
		Initialize $U$ as the list of qubits not measured in $m$

        Sort $U$ in descending order according to $CQ^\tau$
		
		Remove from $U$ the indices where $PSi$ and $PSj$ coincide

        $B$ = [$\mathcal{X}$, $\mathcal{Y}$, $\mathcal{Z}$, $\mathrm{Bell}$, $\Omega^X$, $\Omega^Y$, $\Omega^Z$, $\chi$, $\tilde{\chi}$]

        Sort $B$ in descending order according to $CM^\tau$
		
		\While{Length($U$) $\neq$ 0\label{line:while}}{
			\For{$\varepsilon$ in $B$\label{line:OrdMeas}}{
                
				\For{$p$ in permutations of $U$ of size equal to the number of qubits where $\varepsilon$ acts \label{line:qubitord}}{

                        \If{$\varepsilon$ acts on one qubit}{
                    
						\If{$PSi[p]$ and $PSj[p]$ are compatible with $\varepsilon$}{
							Append [$\varepsilon$, $p$] to $\tilde{m}$
       
							Remove qubit in $p$ from $U$
							
							Go to line \ref{line:while}
						}
					}
                
					\ElseIf{$\tau[p]$ connected in $G$\label{line:processor embedding}}{
                    
						\If{$PSi[p]$ and $PSj[p]$ are compatible with $\varepsilon$}{
							Append [$\varepsilon$, $p$] to $\tilde{m}$
       
							Remove qubits in $p$ from $U$
							
							Go to line \ref{line:while}
						}
					}
				}
			}
			\Return{\textbf{False}, $m$}
		}
		\Return{\textbf{True}, $\tilde m$}
		
	\end{subroutine}

 \begin{figure}[H]
		\centering
		\includegraphics[width=0.8\linewidth]{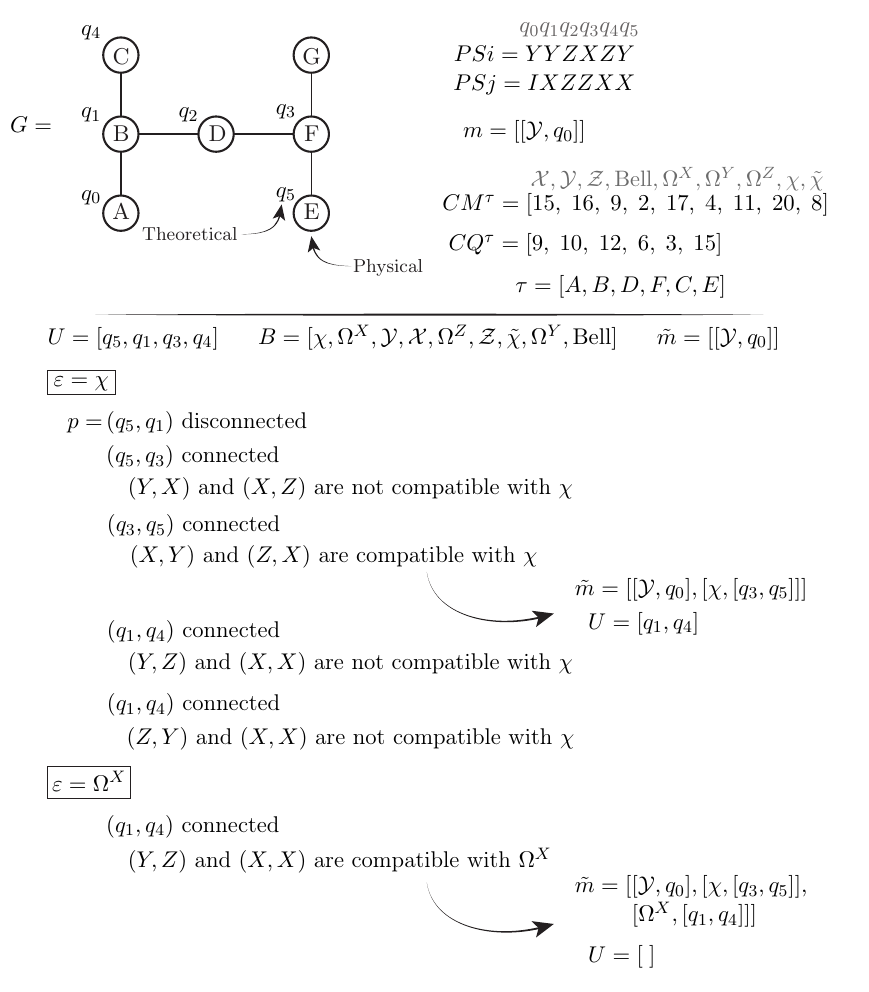}
		\caption{\textbf{Example of Subroutine 2 (Greedy measurement assignment taking chip's connectivity into account).} In this example, the algorithm has to either find HEEMs that can simultaneously measure the Pauli strings $YYZXZY$ and $IXZZXX$ (assuming that the theoretical qubit $q_0$ has been assigned the measurement $\mathcal{Y}$) or otherwise state the impossibility of the task. First of all, the algorithm creates $U$, which is the list of theoretical qubits where there is no assigned measurement nor the Pauli strings coincide. Note that $U$ is sorted in descending order of $CQ^\tau$. $B$ is the vector of possible measurements, sorted in descending order of $CM^\tau$. The first measurement to check is $\chi$. The qubits $q_1$ and $q_5$ are not connected, so we try to measure the qubits $q_5$ and $q_3$, which in fact are compatible with $\chi$. Then the algorithm tries to measure the qubits $q_1$ and $q_4$, but they are not compatible with $\chi$. Thus, the algorithm checks if they are compatible with the \emph{second preferred}  measurement according to $CM^\tau$, which is $\Omega^X$. It succeeds in doing that, so both Pauli strings are compatible with HEEMs. Thus, the subroutine succeeds in finding HEEMs compatible with $PSj$, $PSi$, and the previous measurement defined in $m$.}
\end{figure}

    \vfill
    \newpage
    \subsubsection{Building the processor embedding map}\label{sec:AlgOrd}
	
    In this section, we provide heuristic subroutines to build the \textit{processor embedding} map (Subroutines~\ref{sub:3theophysNoncon} and \ref{sub:4theophysConnected}) and the compatibility matrix (Subroutine~\ref{sub:5CompMat}).
    
    \begin{subroutine}[h]
		\caption{Processor embedding map construction, without ensuring that $\tau(G')$ is connected}\label{sub:3theophysNoncon}
		
		\KwIn{$n$ Pauli strings PS of $N$ qubits the chip's connectivity $G$}
        C = Subroutine~\ref{sub:5CompMat}(PS) \space \textcolor{blue}{// Compatibility matrix}
        
		$AQ$ = [ ]  \textcolor{blue}{// Assigned qubits}
		
		Initialize $\tau$ as a list of size $N$ \space \textcolor{blue}{// Embedding map}
		
		\While{Length($AQ$) $\neq$ $N$}{
			Choose $i$, $j$ such that $C[i,j]$ = Max($C$)
			
			\If{$i$ in $AQ$ and $j$ in $AQ$}{
                \If{$j$ in $AQ$}{
                    Swap $i\leftrightarrow j$ \space \textcolor{blue}{// Ensure that $i\in AQ$ and $j \notin AQ$}
                }
				
				Let $J$ be an unassigned neighbor of $\tau[i]$ in $G$
				
				$\tau[j]=J$
				
				Append $j$ to $AQ$

			}
			\ElseIf{($i$ and $j$) not in $AQ$}{
				Choose physical qubits $I$ and $J$ that are connected in $G$ and unassigned
    
				$\tau[i]$ = $I$
    
                $\tau[j]$ = $J$

                Append $i$ and $j$ to $AQ$
			}

            RemoveAssignedQubits($C$, $\tau[i]$, $AQ$, $G$, $\tau$) \space \textcolor{blue}{// Subroutine~\ref{sub:RemoveAssigned}}
    
            RemoveAssignedQubits($C$, $\tau[j]$, $AQ$, $G$, $\tau$) \space \textcolor{blue}{// Subroutine~\ref{sub:RemoveAssigned}}

            $C[i, j]$ = $C[j, i]$ = NaN
        
		}
		\Return $\tau$ 
	\end{subroutine}

	\begin{subroutine}[H]
		\caption{Processor embedding map construction, ensuring that $\tau(G')$ is connected}\label{sub:4theophysConnected}
		\KwIn{$n$ Pauli strings PS of $N$ qubits and chip's connectivity $G$}
        $C$ = Subroutine~\ref{sub:5CompMat}(PS) \space \textcolor{blue}{// Compatibility matrix}

        $AQ$ = [ ] \space \textcolor{blue}{// Assigned qubits}

		Initialize $\tau$ as a list of size $N$ \space \textcolor{blue}{// Embedding map}
		
		Choose $i$, $j$ such that $C[i,j]$ = Max($C$)
		
		Choose $I$, $J$ a pair of connected qubits in $G$
		
		$\tau[i]$ = $I$

        $\tau[j]$ = $J$

        Append $i$ and $j$ to $AQ$
		
		$C[i,j]$ = $C[j,i]$ = NaN
                
		RemoveAssignedQubits($C$, $\tau[i]$, $AQ$, $G$, $\tau$)  \space \textcolor{blue}{// Subroutine \ref{sub:RemoveAssigned}}
  
		\While{Length($AQ$) $\neq$ $N$}{
			$C'$ = $C[AQ,:]$
			
			Choose $i$, $j$ such that $C'[i,j]$ = Max($C'$)
			
			\If{$j$ not in $AQ$}{
				
				Let $J$ be an unassigned neighbor of $\tau[i]$ in $G$
				
				$\tau[j]$ = $J$
				
                Append $j$ to $AQ$
    
				RemoveAssignedQubits($C$, $\tau[j]$, $AQ$, $G$, $\tau$) \space \textcolor{blue}{// Subroutine \ref{sub:RemoveAssigned}}
			}

            $C[i, j]$ = $C[j, i]$ = NaN
            
		}
		\Return $\tau$ 
		
	\end{subroutine}
 
    \clearpage
    \newpage
    \begin{subroutine}[H]
		\caption{Remove qubits with no free neighbors}\label{sub:RemoveAssigned}
		\KwIn{Compatibility matrix $C$, physical qubit $I$, assigned qubits $AQ$, chip's connectivity $G$ and \textit{processor embedding} map $\tau$}

        Let $PQs$ be the list of assigned neighbors of $I$ in $G$

        Append $I$ to $PQs$
        
		\For{$S$ in $PQs$}{
            Let $s$ be the theoretical qubit assigned to $S$ in $\tau$

            \If{all neighbors of $S$ in $G$ are assigned}{
                    $C[s, :]$ = $C[:, s]$ = NaN
            }
        }
	\end{subroutine}

    \begin{subroutine}[H]	
		\caption{Construction of compatibility matrix}\label{sub:5CompMat}
		\KwIn{$n$ Pauli strings $PS$ of $N$ qubits}
		Initialize $C$ as an $N\times N$ matrix of zeros

        Set diagonal entries of $C$ equal to NaN

        \For{$i$ in [$0$, $\dots$, $N-2$]}{
            \For{$j$ in [$i+1$, $\dots$, $N-1$]}{
            
                $nII$ = 0
                
                \For{$ps$ in $PS$}{
                    $nII$ += ($ps[i,j]$ == $II$) \space \textcolor{blue}{// 1 if True, 0 if False}
                }

                \textcolor{blue}{// Same for $nXX$, $nYY$, $\ldots$}
                
                $n\mathrm{Bell}$ = Binomial($nII+nXX+nYY+nZZ$, 2)
                
                $n\Omega^X$ = Binomial($nII+nXX+nYZ+nZY$, 2)

                \textcolor{blue}{// Repeat for all other measurements}

                .

                .

                .
                
                $C[i,j]$ = $C[j,i]$ = $n\mathrm{Bell}+n\Omega^X +n\Omega^Y+n\Omega^Z+n\chi+n\tilde{\chi}$
    		}
        }
        
		\Return $C$
    
	\end{subroutine}

    \begin{figure}[ht!]
		\centering
		\includegraphics[width=0.85\linewidth]{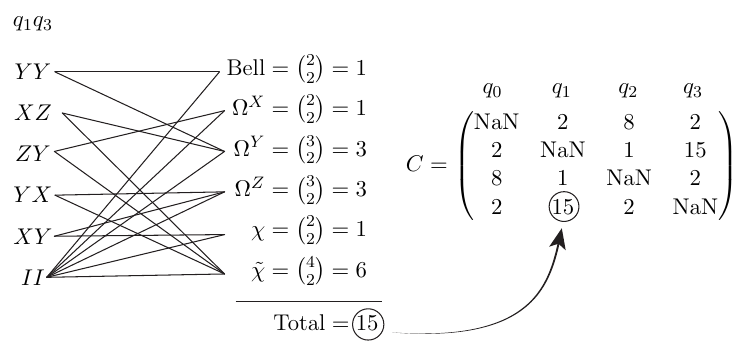}
		\caption{\textbf{Example of Subroutine~\ref{sub:5CompMat} (Construction of compatibility matrix).} The example shows how to compute the entry of the compatibility matrix corresponding to the compatibilities between $q_1$ and $q_3$ (these sub Pauli strings are those that come from the example of Figure~\ref{fig:subroutine_1}). The graph on the left has an edge between a size 2 Pauli string and a HEEM if that Pauli string can be measured with that HEEM. Hence, the number of compatibilities due to the $\tilde{\chi}$ measurement is ${4 \choose 2}$ because $\tilde\chi$ is compatible with 4 Pauli strings and one can make ${4 \choose 2}$ different sets out of $4$ different elements. The analog is true for the rest of the HEEMs. Summing all these numbers gives the total number of compatibilities between the theoretical qubits $q_1$ and $q_3.$}
		\label{fig:subroutine_5}
	\end{figure}
    
    \newpage
    \subsubsection{Choosing the iterative orders} 
    Finally, Subroutine~\ref{sub:6tauCompMat} outputs the $\tau$-compatibility matrix $C^\tau$ and the vectors $CM^\tau$ and $CQ^\tau$, which determines the order in which Subroutine~\ref{sub:2assignMeas} runs through the iterative elements. 
    
    \begin{subroutine}[H]
		\caption{Choosing the iterative orders}\label{sub:6tauCompMat}
		
		\KwIn{$n$ Pauli strings $PS$ of $N$ qubits, \textit{processor embedding} map $\tau$ and chip's connectivity $G$}
		
		Initialize $C^\tau$ as an $N\times N$ matrix of zeros \space \textcolor{blue}{// Compatibility matrix}

        Set diagonal entries of $C^\tau$ equal to NaN

        Initialize $CM^\tau$ as a list of size 9 with zeros \space \textcolor{blue}{// Measurement compabilities}

        Initialize $CQ^\tau$ as a list of size N with zeros \space \textcolor{blue}{// Qubit compabilities}
		
        Identify $\mathcal{X}=0$, $\mathcal{Y}=1$, $\mathcal{Z}=2$, $\mathrm{Bell} = 3$, $\Omega^X = 4$, $\ldots$  \space \textcolor{blue}{// Analogue for all measurements}
  
		\For{$i$ in [$0$, $\dots$, $N-1$]}{           

            $nI$ = 0
                
            \For{$ps$ in $PS$}{
                $nI$ += ($ps[i]$ == $I$) \space \textcolor{blue}{// 1 if True, 0 if False}
            }

            \textcolor{blue}{// Same for $nX$, $nY$ and $nZ$}
            
			$CQ^\tau[i]$ += Binomial($nI+nX$, 2) + Binomial($nI+nY$, 2) + Binomial($nI+nZ$, 2)
				
			$CM^\tau[\mathcal{X}]$ += Binomial($nI+nX$, 2)
   
            $CM^\tau[\mathcal{Y}]$ += Binomial($nI+nY$, 2)

            $CM^\tau[\mathcal{Z}]$ += Binomial($nI+nZ$, 2)
				
			\For{$j$ in [$i+1$, $\dots$, $N-1$]}{
				\If{$\tau[i]$ and $\tau[j]$ connected in $G$}{
                    Define $nXX$, $nXY$, $nXZ$, $\ldots$, similar to $nI$ for $ps[i,j]$
          
					$n\mathrm{Bell}$ = Binomial($nII+nXX+nYY+nZZ$, 2)
     
                    $n\Omega^X$ = Binomial($nII+nXX+nYZ+nZY$, 2) 

                    \textcolor{blue}{// Repeat for all other measurements}

                    .

                    .

                    .
					
					$C^\tau [i,j]$ = $C^\tau [j,i]$ = $n\mathrm{Bell}+n\Omega^X+n\Omega^Y+n\Omega^Z+n\chi+n\tilde{\chi}$
					
					$CM^\tau[\mathrm{Bell}]$ += $n\mathrm{Bell}$
     
                    $CM^\tau[\Omega^X]$ += $n\Omega^X$ 

                    .

                    .

                    .
                    
                    $CQ^\tau[i]$ += $C^\tau[i,j]$
                    
                    $CQ^\tau[j]$ += $C^\tau[i,j]$
					
				}
			}
            
		}
  
		\Return $C^\tau$, $CM^\tau$, $CQ^\tau$
		
	\end{subroutine}

    \newpage
    \begin{figure}[H]
		\centering
		\includegraphics[width=0.85\linewidth]{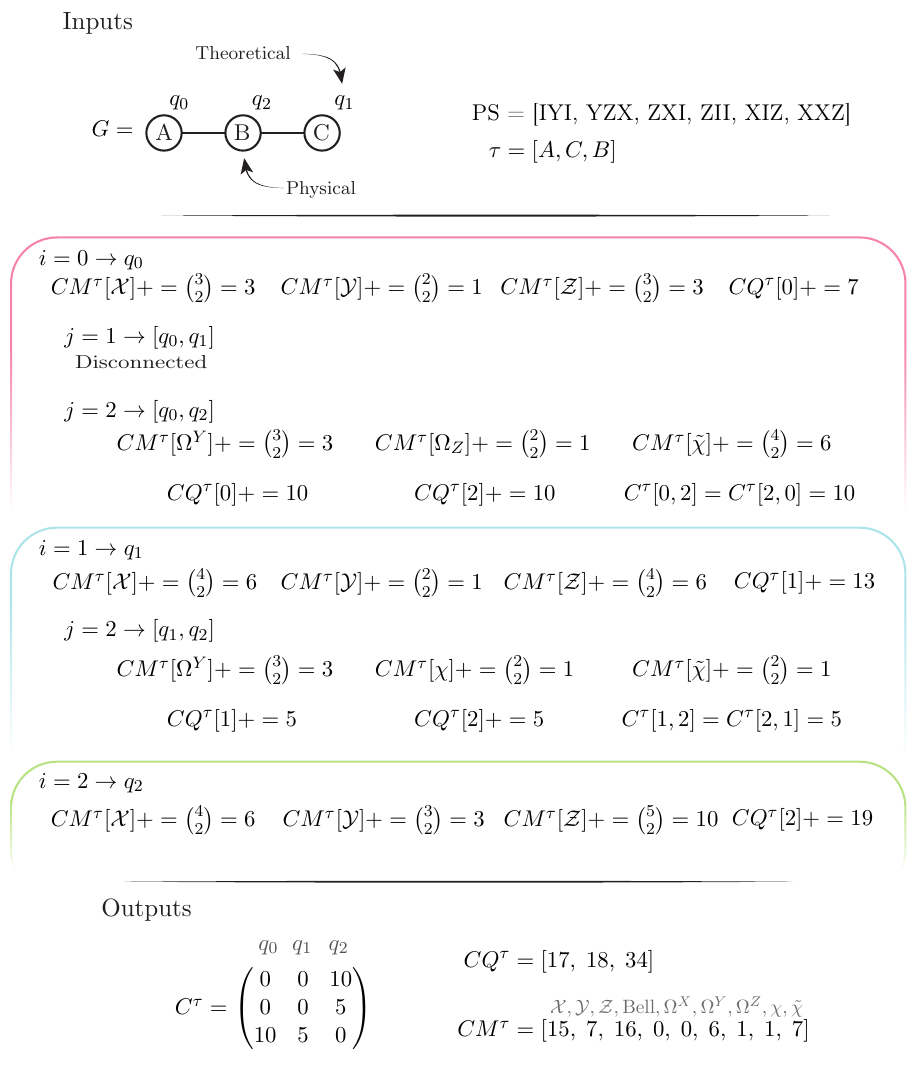}
		\caption{\textbf{Example of Subroutine 6 (Choosing the iterative orders).} The subroutine inputs a list of Pauli strings $PS$ and a \textit{processor embedding} map $\tau$, and chooses an adequate way of running through the qubits' and measurements' loops of Subroutine~\ref{sub:2assignMeas}. To do that, it computes the vectors $CM^\tau$ and $CQ^\tau$, respectively. It begins by choosing the first theoretical qubit, $q_0$. Then it computes how many one-qubit $\mathcal{X}$-compatibilities arise in that qubit and adds them to $CM^\tau[\mathcal{X}]$ and to $CQ^\tau[0]$. The same is done for $\mathcal{Y}$ and $\mathcal{Z}$. Then, it runs through all pairs of theoretical qubits including $q_0$. First, it checks if $\tau$ maps the pair of qubits to connected physical qubits. If so, it counts the compatibilities due to two-qubit measurements involving that pair and adds them to the corresponding entries of $C^\tau$, $CM^\tau$, and $CQ^\tau$. After that, the same is done with the remaining qubits.}
		\label{fig:subroutine_6}
	\end{figure}

    \newpage
    \section{Asymptotic scaling} \label{SMsec:scaling}
\begin{figure}[htb!]
    	\centering
    	\includegraphics[width=0.85\linewidth]{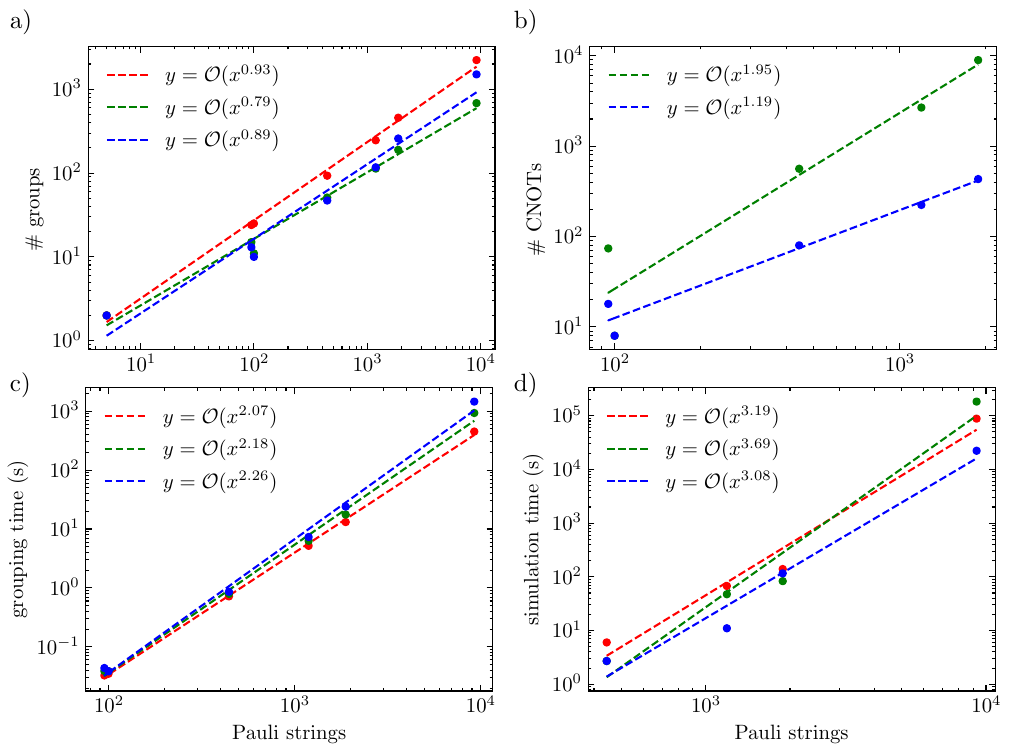}
    	\caption{Dependence of \textbf{a)} the total number of groups, \textbf{b)} the number of CNOTs gates, \textbf{c)} CPU time for the grouping, and \textbf{d)} CPU time for the simulation of different molecules on the number of Pauli strings. The grouping algorithms used are TPB (red), EM (green), and HEEM (blue). The dashed lines represent a fit to the function $y = \beta x^\alpha$. Both EM and HEEM use \textit{ibmq\_montreal} connectivity. In \textbf{d)}, the simulation mimics the architecture, base gate, and noise of the quantum device \textit{ibmq\_montreal}. Each simulation has a total of $2^{14}$ shots evenly distributed across all measurements in each grouping. The simulation is performed using the qasm HPC simulator provided by IBMQ. The transpiling and queue times are not included in these times.}
    	\label{fig:scalings}
    \end{figure}

    In Fig.~\ref{fig:scalings} a), we show the dependence of the number of groups on the total number of Pauli strings in the Hamiltonian of different molecules. We found that TPB obtains the largest number of groups for all molecules, while EM results in a significant reduction in the number of groups. HEEM obtains better results than TPB, but due to the constraint in the non-connected measurements, it does not reach the results of EM. However, in Fig.~\ref{fig:scalings} b), we find a considerable improvement in the total number of CNOT gates. After transpiling, taking into account a real quantum device architecture, EM needs more CNOT gates to perform SWAP gates between non-connected qubits. This problem is solved by using HEEM, which does not need any SWAP gate, resulting in fewer CNOT gates. 
    
     It is also important to consider the time needed to perform the grouping on a classical CPU, see Fig.~\ref{fig:scalings} c). All three algorithms begin with the construction of the Pauli graph, which contains information about the commutative Pauli strings. Once the graph is obtained, TPB uses LDFC for graph coloring, whose time complexity is $\order{n^2}$, where $n$ is the total number of Pauli strings. EM uses a more sophisticated algorithm which needs to run through the graph and check if it can group a pair of terms with any of the existing bases. This extra check results in a slower algorithm. Furthermore, HEEM also checks if the grouping is compatible with the chip's connectivity. However, despite the fact that the HEEM grouping is slower than the other methods, looking at the simulation time, Fig.~\ref{fig:scalings} d), it is clear that the reduction in the simulation time is notorious. Even if TPB circuits do not require CNOT gates, the number of circuits to simulate is much larger than that of HEEM. On the other hand, EM grouping results in lower circuits, but the number of SWAP gates needed to perform the simulation grows much faster than that of HEEM, making it the slower algorithm to simulate. 

     \section{Variational circuits}\label{variational_circuit}
     To compute the VQE of the H$_2$O molecule given in the main text, we use a circuit similar to the one shown below. First, the circuit is initialized with a Hartree-Fock state. The variational ansatz used is known as EfficientSU2, which is composed of a first layer of one-qubit rotations around the $y$-axis with the angles $\theta_i$, and a second layer of one-qubit rotations around the $z$-axis. Then the entangling layer is composed of CNOT gates for all well-connected qubits on the chip. After that, two more one-qubit rotations around the $y$- and $z$-axes are applied. There is a total of $4N$ degrees of freedom, where $N$ is the total number of qubits. Finally, we apply different one- and two-qubit gates to group the Hamiltonian terms, as explained in the main text, and measure the qubits.
     
     \begin{center}
         \begin{quantikz}
		\lstick{$q_0$}&\gate[6,nwires={5},disable auto height]{\begin{array}{c} \text{H} \\ \text{A} \\\text{R} \\\text{T} \\\text{R} \\\text{E} \\\text{E} \\ \\ \text{F} \\ \text{O} \\ \text{C} \\ \text{K} \end{array}} \slice{initialize} &\gate{R_Y(\theta_0)} & \gate{R_Z(\theta_{N})} & \ctrl{1}\gategroup[6,steps=3,style={dashed,rounded corners,fill=blue!15, inner xsep=2pt}, background]{{\sc well-connected}} & \qw & \qw & \gate{R_Y(\theta_{2N})} & \gate{R_Z(\theta_{3N})} & \gate[6,nwires={5},disable auto height]{\begin{array}{c} \text{G} \\ \text{R} \\\text{O} \\\text{U} \\\text{P} \\\text{I} \\\text{N} \\ \text{G}\end{array}} & \meter{} \\
		
		\lstick{$q_1$}& &\gate{R_Y(\theta_1)} & \gate{R_Z(\theta_{N + 1})} & \targ{} & \ctrl{1} & \ctrl{2} & \gate{R_Y(\theta_{2N + 1})} & \gate{R_Z(\theta_{3N + 1})} \slice{ansatz} & & \meter{}\\
		
		\lstick{$q_2$}& &\gate{R_Y(\theta_2)} & \gate{R_Z(\theta_{N + 2})} & \qw & \targ{} & \qw & \gate{R_Y(\theta_{2N + 2})} & \gate{R_Z(\theta_{3N + 2})} & & \meter{} \\
		
		\lstick{$q_3$}& &\gate{R_Y(\theta_3)} & \gate{R_Z(\theta_{N + 3})} & \ctrl{2} & \qw & \targ{} & \gate{R_Y(\theta_{2N +3})} & \gate{R_Z(\theta_{3N +3})} & & \meter{}\\
		
		\lstick{$\vdots$} &  & \vdots & \vdots & & & & \vdots & \vdots & & \vdots\\
		
		\lstick{$q_{N-1}$} & &\gate{R_Y(\theta_{N - 1})} & \gate{R_Z(\theta_{2N - 1})} & \targ{} & \qw & \qw & \gate{R_Y(\theta_{3N - 1})} &\gate{R_Z(\theta_{4N - 1})} & & \meter{}
	\end{quantikz}
        \begin{figure}[H]
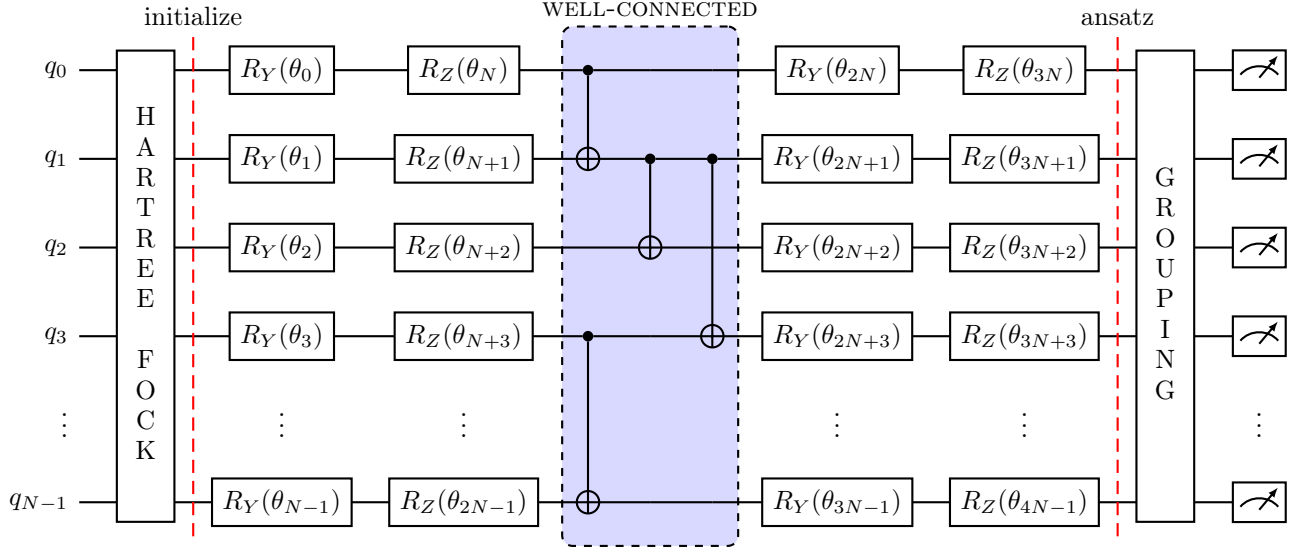

            \caption{Variational circuit for variational quantum eigensolver with a total of $N$ qubits, and $4N$ rotation angles $\theta_i$. The entangling gates are only performed between qubits that are connected in the quantum chip.}
            \label{fig:var_circ}
        \end{figure}
     \end{center}
     
\end{document}